%% file: racah.tex
\documentclass[aps,prb,showpacs,floatfix,superscriptaddress,tightenlines,12pt]{revtex4}

\usepackage{graphicx}
\usepackage{amsmath}
\usepackage{amssymb}
\usepackage{mathtools}
\usepackage{liealg}

\DeclareRobustCommand{\abs}[1]{\lvert#1\rvert}
\renewcommand{\vec}[1]{\boldsymbol{\mathbf{\mathrm{#1}}}}
\DeclareRobustCommand{\subqn}[2]{\underset{#2}{#1}}
\DeclareRobustCommand{\subqnsupset}[1]{\underset{#1}{\supset\mathrel{\makebox[0pt]{\phantom{()}}}}}

\DeclareRobustCommand{\half}{\tfrac12}
\DeclareRobustCommand{\shalf}{1/2}
\DeclareRobustCommand{\chain}[1]{\ensuremath{\mathrm{#1}}}

\DeclareRobustCommand{\Jhat}{\hat{J}}
\DeclareRobustCommand{\Thh}{T^{(\half\half)}}
\DeclareRobustCommand{\Tss}{T^{(1/2\,1/2)}}   
\DeclareRobustCommand{\Tfull}[2]{\Thh_{#1\half#2\half}}
\DeclareRobustCommand{\Tpp}{T_{++}}
\DeclareRobustCommand{\Tpm}{T_{+-}}
\DeclareRobustCommand{\Tmp}{T_{-+}}
\DeclareRobustCommand{\Tmm}{T_{--}}
\DeclareRobustCommand{\scrM}{\mathcal{M}}
\DeclareRobustCommand{\scrN}{\mathcal{N}}
\DeclareRobustCommand{\bbR}{\mathbb{R}}
\DeclareMathOperator{\mult}{mult}

\DeclareRobustCommand{\ad}{a^\dagger}
\DeclareRobustCommand{\at}{\tilde{a}}

\DeclareRobustCommand{\jhat}{\hat{\jmath}}

\makeatletter
\def\hline@rule{%
 \hrule \@height \arrayrulewidth
}%
\makeatother

\makeatletter
\DeclareRobustCommand{\methreefour}[3]{
\LIEALG@dell{1}{4}{\langle}
\begin{LIEALG@matrix}{1}{3} #1\end{LIEALG@matrix}
\,\LIEALG@vert{1}{4}\,
\begin{LIEALG@matrix}{1}{4} #2\end{LIEALG@matrix}
\,\LIEALG@vert{1}{4}\,
\begin{LIEALG@matrix}{1}{4} #3\end{LIEALG@matrix}
\LIEALG@delr{1}{3}{\rangle}
}
\makeatother

\begin{document}


\title{\boldmath Racah's method for general subalgebra chains: \\ Coupling
coefficients of $\grpso{5}$ in canonical and physical bases}

\author{M. A. Caprio}
\affiliation{Department of Physics, University of Notre Dame,
Notre Dame, Indiana 46556-5670, USA}
\author{K. D. Sviratcheva}
\affiliation{Department of Physics and Astronomy, 
Louisiana State University, Baton Rouge, Louisiana 70803-4001, USA}
\author{A. E. McCoy}
\affiliation{Department of Physics, University of Notre Dame,
Notre Dame, Indiana 46556-5670, USA}
\affiliation{Department of Physics, Grinnell College, 
Grinnell, Iowa 50112-1690, USA}

\date{\today}

\begin{abstract}
It is shown that the method of infinitesimal generators (``Racah's
method'') can be broadly and systematically formulated as a method
applicable to the calculation of reduced coupling coefficients for a
generic subalgebra chain $G\supset H$, provided the reduced matrix
elements of the generators of $G$ and the recoupling coefficients of
$H$ are known.  The calculation of $\grpso{5}\supset\grpso{4}$ reduced
coupling coefficients is considered as an example, and a procedure for
transformation of reduced coupling coefficients between canonical and
physical subalegebra chains is presented.  The problem of calculating
coupling coefficients for generic irreps of $\grpso{5}$, reduced with
respect to any of its subalgebra chains, is completely resolved by
this approach.
\end{abstract}


\maketitle

\section{Introduction}
\label{sec-intro}

Continuous symmetries and their associated Lie algebras facilitate the
description of many-body systems both directly and indirectly.  When a
symmetry occurs as a dynamical symmetry of the system, the
corresponding algebra immediately gives the spectroscopic properties
of the system.  However, even when a symmetry is strongly broken, the
algebraic structure nonetheless provides a calculational tool,
classifying the basis states used in a full computational treatment of
the many-body problem and greatly simplifying the underlying
calculational machinery. Lie algebras have a long history of
application, in both these capacities, to nuclear spectroscopy and
related
problems.\cite{racah1965-group-spectro,hecht1973:nuclear-symmetries,iachello2006:liealg}
The fundamental quantities underlying calculations within a Lie
algebraic framework are the coupling coefficients of the algebra, also
known as generalized Clebsch-Gordan coefficients or Wigner
coefficients.  These are needed in order to couple states (or
operators) of good symmetry to yield new states (or operators) of good
symmetry, and they are required for the calculation of matrix elements
through the generalized Wigner-Eckart theorem of the algebra.

The Lie algebra $\grpso{5}$, isomorphic to $\grpsp{4}$, has several
distinct applications in nuclear theory, involving different physical
realizations of the operators, and in which different subalgebra
chains are relevant to the symmetry properties.  The natural
construction of $\grpso{5}$ in terms of generators of rotation in
five-dimensional space gives rise to a canonical
$\grpso{4}\sim\grpso{3}\otimes\grpso{3}$ subalgebra.\cite{fn-apologia}
However, application as the proton-neutron pairing quasispin
algebra\cite{helmers1961:shell-sp,flowers1964:quasispin,ichimura1965:seniority-isospin,ginocchio1965:so5-quasispin,hecht1967:so5-shell-wigner}
requires reduction with respect to the $\grpu{1}\otimes\grpso{3}$
algebra of isospin and occupation number operators.  For the dynamics
of spin-$2$ bosons (as in the interacting boson
model\cite{arima1976:ibm-u5,iachello1987:ibm}) and for the Bohr
collective
model,\cite{bohr1998:v2,eisenberg1987:v1,rowe2004:spherical-harmonics,debaerdemacker2008:collective-cartan,rowe2009:acm}
the appropriate reduction is instead with respect to a physical
angular momentum $\grpso{3}$ subalgebra.

In this article, it is shown that the method of infinitesimal
generators (``Racah's method'') can be broadly and systematically
formulated as a method applicable to the calculation of reduced
coupling coefficients for a generic subalgebra chain $G\supset H$,
provided the reduced matrix elements of the generators of $G$ and the
recoupling coefficients of $H$ are known (Sec.~\ref{sec-method}).  More specifically, the
problem of calculating coupling coefficients for generic irreps of
$\grpso{5}$, reduced with respect to any of the subalgebra chains, is
completely resolved by this approach.  The calculation of reduced
coupling coefficients for the $\grpso{5}\supset\grpso{4}$ canonical
chain is considered in detail (Sec.~\ref{sec-so5}).  Coupling coefficients reduced with
respect to the noncanonical subalgebra chains of $\grpso{5}$ may be
obtained by a similar application of Racah's method, or they can be
deduced from the canonical chain coupling coefficients by unitary
transformation.  The general formulation in the presence of outer
multiplicities for $H$, numerical examples for $\grpso{5}$, and a
detailed account of the transformation procedure between subalgebra
chains are given in the appendices.

\section{Method}
\label{sec-method}

\subsection{Background and definitions}
\label{sec-background}

Consider a Lie algebra $G$ and subalgebra $H$.  States which reduce
this subalgebra chain may be identified by the irrep labels $\Gamma$
of $G$, the irrep labels $\Lambda$ of $H$, and a label $\lambda$
(typically the Cartan weights) to distinguish basis states within
$\Lambda$, as $\ket{\Gamma\Lambda\lambda}$.  The coupling
coefficients, or generalized Clebsch-Gordan coefficients, for $G$
relate the uncoupled product states of two irreps of $G$ to the
coupled states, as
\begin{equation}
\label{eqn-coupling-simple}
\ket[4]{\Gamma_1~~\Gamma_2\\\Gamma\\\Lambda\\\lambda}
=
\sum_{\substack{\Lambda_1\Lambda_2\\\lambda_1\lambda_2}}
\ccg{\Gamma_1}{\Lambda_1}{\lambda_1}{\Gamma_2}{\Lambda_2}{\lambda_2}{\Gamma}{\Lambda}{\lambda}
\ket[3]{\Gamma_1&\Gamma_2\\\Lambda_1&\Lambda_2\\\lambda_1&\lambda_2}
\end{equation}
In general, additional labels will be required to resolve
multiplicities.  There may be ``outer'' multiplicities in the
Clebsch-Gordan series for the outer product of $G$ (that is,
$\Gamma_1\otimes\Gamma_2$ may contain the irrep $\Gamma$ more than
once), and there may be ``branching'' multiplicites under the
restriction of $G$ to $H$ (that is, the given irrep $\Gamma$ of $G$
may contain an irrep $\Lambda$ of $H$ more than once).  The coupling
relation~(\ref{eqn-coupling-simple}) generalizes, with multiplicities,
to
\begin{equation}
\label{eqn-coupling-mult}
\ket[4]{\Gamma_1~~\Gamma_2\\\rho\Gamma\\a\Lambda\\\lambda}
=
\sum_{\substack{a_1\Lambda_1a_2\Lambda_2\\\lambda_1\lambda_2}}
\ccg{\Gamma_1}{a_1\Lambda_1}{\lambda_1}{\Gamma_2}{a_2\Lambda_2}{\lambda_2}{\rho\Gamma}{a\Lambda}{\lambda}
\ket[3]{\Gamma_1&\Gamma_2\\a_1\Lambda_1&a_2\Lambda_2\\\lambda_1&\lambda_2},
\end{equation}
where $\rho$ is the outer multiplicity index for $G\otimes
G\rightarrow G$, and the $a$ indices resolve the branching
multiplicities for $G\rightarrow H$.  Furthermore, $H$ may be subject
to outer multiplicites ($H\otimes H \rightarrow H$).  In the following
discussion, we shall for simplicity take the subalgebra $H$ to be
multiplicity free.  Such is the case for the commonly encountered
situation in which the physically relevant subalgebra $H$ is
$\grpso{3}$, as well as for the subalgebra $\grpso{4}$ considered in
Sec.~\ref{sec-so5}.  However, the necessary generalizations in the
presence of outer multiplicities on $H$ are given in
Appendix~\ref{app-mult}, as would be needed for consideration of, {\it
e.g.}, chains involving $\grpsu{3}$ as a subalgebra.

Racah's factorization
lemma\cite{racah1949:complex-spectra-part4-f-shell} allows the
coupling coefficient appearing in~(\ref{eqn-coupling-mult}) to be
decomposed as the product
\begin{equation}
\label{eqn-factorization}
\ccg{\Gamma_1}{a_1\Lambda_1}{\lambda_1}{\Gamma_2}{a_2\Lambda_2}{\lambda_2}{\rho\Gamma}{a\Lambda}{\lambda}
=
\cg{\Lambda_1}{\lambda_1}{\Lambda_2}{\lambda_2}{\Lambda}{\lambda}
\cg{\Gamma_1}{a_1\Lambda_1}{\Gamma_2}{a_2\Lambda_2}{\rho\Gamma}{a\Lambda}
\end{equation}
of a coupling coefficient of $H$, embodying all the dependence upon
weights $\lambda$, with a {\it reduced coupling coefficient} (or {\it
isoscalar factor}) for $G\supset H$.  The reduced coupling coefficient is
nonzero only if $\Gamma$ is contained in the outer product of
$\Gamma_1$ and $\Gamma_2$ ({\it i.e.}, $\Gamma_1\otimes
\Gamma_2\rightarrow \Gamma$), each irrep of $H$ is contained in the
corresponding irrep of $G$ ({\it i.e.},
$\Gamma_1\rightarrow\Lambda_1$, $\Gamma_2\rightarrow\Lambda_2$, and
$\Gamma\rightarrow\Lambda$), and $\Lambda$ is contained in the outer
product of $\Lambda_1$ and $\Lambda_2$ ({\it i.e.}, $\Lambda_1\otimes
\Lambda_2\rightarrow \Lambda$).  The reduced coupling coefficients
satisfy the orthonormality conditions\cite{wybourne1974:groups}
\begin{align}
\label{eqn-ortho-bra-sum}
\sum_{a_1\Lambda_1a_2\Lambda_2}
\cg{\Gamma_1}{a_1\Lambda_1}{\Gamma_2}{a_2\Lambda_2}{\rho\Gamma}{a\Lambda}
\cg{\Gamma_1}{a_1\Lambda_1}{\Gamma_2}{a_2\Lambda_2}{\rho'\Gamma'}{a'\Lambda}
&=\delta_{(\rho\Gamma)(\rho'\Gamma')}\delta_{aa'}
\intertext{and}
\label{eqn-ortho-ket-sum}
\sum_{\rho\Gamma a}
\cg{\Gamma_1}{a_1\Lambda_1}{\Gamma_2}{a_2\Lambda_2}{\rho\Gamma}{a\Lambda}
\cg{\Gamma_1}{a_1'\Lambda_1'}{\Gamma_2}{a_2'\Lambda_2'}{\rho\Gamma}{a\Lambda}
&=\delta_{(a_1\Lambda_1)(a_1'\Lambda_1')}\delta_{(a_2\Lambda_2)(a_2'\Lambda_2')},
\end{align}
for any irrep $\Lambda$ such that $\Gamma\rightarrow\Lambda$.

If $T^{\Lambda_T}$ is an irreducible tensor operator with respect to
$H$, the Wigner-Eckart theorem for $H$ permits the expression of a
general matrix element of $T^{\Lambda_T}_{\lambda_T}$
as\cite{wybourne1974:groups}
\begin{equation}
\label{eqn-we}
\me[3]{\Gamma'\\a'\Lambda'\\\lambda'}{T^{\Lambda_T}_{\lambda_T}}{\Gamma\\a\Lambda\\\lambda}
=
\cg{\Lambda}{\lambda}{\Lambda_T}{\lambda_T}{\Lambda'}{\lambda'}
\rme[2]{\Gamma'\\a'\Lambda'}{T^{\Lambda_T}}{\Gamma\\a\Lambda},
\end{equation}
in terms of a coupling coefficient for $H$ and a {\it reduced matrix
element} with respect to $H$.  A Wigner-Eckart theorem of this form
[or its generalization~(\ref{eqn-we-mult})] may be obtained whenever
$H$ is a compact, semi-simple Lie algebra.

Several methods may be
considered, in general, for constructing the reduced coupling
coefficients of Lie algebras:
\begin{enumerate}
\item[(1)] Recurrence relations among coupling coefficients may be obtained
by considering the action of an infinitesimal generator
$G_i=G_i^{(1)}+G_i^{(2)}$ on uncoupled and coupled states.  This
approach, used in the present construction, is broadly termed
``Racah's method'' (see Ref.~\onlinecite{wybourne1974:groups}) and
generalizes the classic recurrence method for evaluating
$\grpsu{2}\sim\grpso{3}$ Clebsch-Gordan
coefficients.\cite{edmonds1960:am}

\item[(2)] Recurrence relations and seed values may be obtained by considering the action of a
``shift tensor'', lying outside the
algebra, which connects different irreps of the algebra.\cite{rowe1997:su3-cg-algebraic}  

\item[(3)] Consistency relations among coupling and recoupling coefficients
serve as the basis for a ``building up''
process,\cite{hecht1967:so5-shell-wigner,hecht1969:su4-wigner} in which unknown coupling coefficients can
be deduced from a few known coefficients.

\item[(4)] Explicit realizations of an algebra can be obtained in terms of
bosonic or fermionic creation and annihilation operators.  Relations among coupling
coefficients follow from considering the matrix elements of tensor
operators acting on bosonic or fermionic states
({\it e.g.}, Ref.~\onlinecite{vanisacker1987:spin6-spin5-isf}).  This
approach is generally restricted to symmetric irreps, antisymmetric irreps, or
irreps which can be obtained as simple combinations thereof.
\end{enumerate}
Indeed, all of these approaches have been applied or suggested, in various forms,
for the calculation of specific classes of $\grpso{5}$ coupling
coefficients.\cite{racah1965-group-spectro,hecht1965:so5-wigner,hecht1967:so5-shell-wigner,hemenger1970:so5-quasispin,wybourne1974:groups,iachello1981:ibfm-spin6,vanisacker1987:spin6-spin5-isf,hecht1989:so5-vcs-cg,hecht1993:so5-u2-vcs,han1993:so5-coupling,caprio2007:geomsuper2}
For the {\it symmetric} irreps of $\grpso{5}$, one may also work with an
explicit realization in terms of five-dimensional spherical
harmonics as functions on the four-sphere.  Their triple
overlap integrals are then proportional to $\grpso{5}$ coupling
coefficients.\cite{rowe2004:spherical-harmonics,caprio2009:gammaharmonic}

\subsection{Racah relations among reduced coupling coefficients}
\label{sec-relations}

Let us now consider how the first approach, {\it i.e.}, Racah's
method\cite{racah1949:complex-spectra-part4-f-shell,racah1965-group-spectro}
based on the action of infinitesimal generators, can be generally and
systematically formulated as a method applicable to the calculation of
reduced coupling coefficients involving generic irreps of an arbitrary
subalgebra chain.  Consider the action of an
infinitesimal generator $G_i$ of $G$ on the coupled product state
of~(\ref{eqn-coupling-mult}).  The generator on the product space is
of the form $G_i=G_i^{(1)}+G_i^{(2)}$, where $G_i^{(1)}$ acts only on
the space carrying the irrep $\Gamma_1$ and $G_i^{(2)}$ acts only on
the space carrying the irrep $\Gamma_2$.  The equivalence of the
action of $G_i$ on the two sides of~(\ref{eqn-coupling-mult}) imposes
conditions on the coupling coefficients connecting the different basis
states used on the two sides.  For effective
application of Racah's method, it is most convenient to recast these
relations among coupling coefficients so that they involve only
(1)~{\it reduced} coupling coefficients of $G$ with respect to $H$,
(2)~{\it reduced} matrix elements of the generators of $G$, and
(3)~recoupling coefficients of $H$, as obtained in this section.

Racah's approach requires that the action of the generators on the
basis states of an irrep be known explicitly.  In general, if coupling
coefficients are to be determined for states which reduce $G\supset
H$, it is necessary to consider the action of the generators which are
in $G$ but not in $H$, since only these generators can connect
different irreps $\Lambda$ of $H$.  Note that Racah's method is
essentially an extension of the classic scheme\cite{edmonds1960:am}
for calculating the ordinary $\grpso{3}$ Clebsch-Gordan coefficients,
via recurrence relations obtained by considering the known matrix
elements of $J_\pm=J_\pm^{(1)}+J_\pm^{(2)}$, between the uncoupled
product states $\smallket[2]{J_1&J_2\\M_1&M_2}$ and coupled states
$\smallket[2]{J\\M}$ (see also Sec.~\ref{sec-solution}).

If $T^{\Lambda_T}_{\lambda_T}$ is a generator of $G$, expressed as an
irreducible tensor operator with respect to $H$, we begin by
considering the matrix element of
$T^{\Lambda_T}_{\lambda_T}=T^{\Lambda_T\,(1)}_{\lambda_T}+T^{\Lambda_T\,(2)}_{\lambda_T}$,
between uncoupled and coupled product states,
\begin{equation}
\label{eqn-racah-raw}
\methreefour{\Gamma_1&\Gamma_2\\a_1\Lambda_1&a_2\Lambda_2\\\lambda_1&\lambda_2}{T^{\Lambda_T}_{\lambda_T}}{\Gamma_1~\Gamma_2\\\rho\Gamma\\a\Lambda\\\lambda}
=
\methreefour{\Gamma_1&\Gamma_2\\a_1\Lambda_1&a_2\Lambda_2\\\lambda_1&\lambda_2}{T^{\Lambda_T\,(1)}_{\lambda_T}}{\Gamma_1~\Gamma_2\\\rho\Gamma\\a\Lambda\\\lambda}
+
\methreefour{\Gamma_1&\Gamma_2\\a_1\Lambda_1&a_2\Lambda_2\\\lambda_1&\lambda_2}{T^{\Lambda_T\,(2)}_{\lambda_T}}{\Gamma_1~\Gamma_2\\\rho\Gamma\\a\Lambda\\\lambda}.
\end{equation}
The coupling relation~(\ref{eqn-coupling-mult}) may be used to express
each ket on the right hand side of~(\ref{eqn-racah-raw}) entirely in
terms of uncoupled states, and its inverse, obtained by orthonormality
of coupling coefficients, may be used to express the bra on the left hand side
entirely in terms of coupled states.  Since $T^{\Lambda_T}$, as a
generator of $G$, does not connect different irreps of $G$, and since
the matrix elements of $T^{\Lambda_T}$ between states within an irrep
of $G$ depends only upon the irrep labels, the result
simplifies to
\begin{multline}
\label{eqn-racah-coupled-1-2}
\sum_{\substack{a'\Lambda'\\(\lambda')}}
\me[3]{\Gamma\\a'\Lambda'\\\lambda'}{T^{\Lambda_T}_{\lambda_T}}{\Gamma\\a\Lambda\\\lambda}
\ccg{\Gamma_1}{a_1\Lambda_1}{\lambda_1}{\Gamma_2}{a_2\Lambda_2}{\lambda_2}{\rho\Gamma}{a'\Lambda'}{\lambda'}
=
\sum_{\substack{a_1'\Lambda_1'\\(\lambda_1')}}
\me[3]{\Gamma_1\\a_1\Lambda_1\\\lambda_1}{T^{\Lambda_T}_{\lambda_T}}{\Gamma_1\\a_1'\Lambda_1'\\\lambda_1'}
\ccg{\Gamma_1}{a_1'\Lambda_1'}{\lambda_1'}{\Gamma_2}{a_2\Lambda_2}{\lambda_2}{\rho\Gamma}{a\Lambda}{\lambda}
\\
+
\sum_{\substack{a_2'\Lambda_2'\\(\lambda_2')}}
\me[3]{\Gamma_2\\a_2\Lambda_2\\\lambda_2}{T^{\Lambda_T}_{\lambda_T}}{\Gamma_2\\a_2'\Lambda_2'\\\lambda_2'}
\ccg{\Gamma_1}{a_1\Lambda_1}{\lambda_1}{\Gamma_2}{a_2'\Lambda_2'}{\lambda_2'}{\rho\Gamma}{a\Lambda}{\lambda}.
\end{multline}

To introduce reduced coupling coefficients and reduced matrix
elements, we apply Racah's factorization lemma~(\ref{eqn-factorization})
and the Wigner-Eckart theorem~(\ref{eqn-we}), yielding
\begin{multline}
\label{eqn-racah-semireduced}
\sum_{\substack{a'\Lambda'\\(\lambda')}}
\cg{\Lambda}{\lambda}{\Lambda_T}{\lambda_T}{\Lambda'}{\lambda'}
\rme[2]{\Gamma\\a'\Lambda'}{T^{\Lambda_T}}{\Gamma\\a\Lambda}
\cg{\Lambda_1}{\lambda_1}{\Lambda_2}{\lambda_2}{\Lambda'}{\lambda'}
\cg{\Gamma_1}{a_1\Lambda_1}{\Gamma_2}{a_2\Lambda_2}{\rho\Gamma}{a'\Lambda'}
\\=
\sum_{\substack{a_1'\Lambda_1'\\(\lambda_1')}}
\cg{\Lambda_1'}{\lambda_1'}{\Lambda_T}{\lambda_T}{\Lambda_1}{\lambda_1}
\rme[2]{\Gamma_1\\a_1\Lambda_1}{T^{\Lambda_T}}{\Gamma_1\\a_1'\Lambda_1'}
\cg{\Lambda_1'}{\lambda_1'}{\Lambda_2}{\lambda_2}{\Lambda}{\lambda}
\cg{\Gamma_1}{a_1'\Lambda_1'}{\Gamma_2}{a_2\Lambda_2}{\rho\Gamma}{a\Lambda}
\\+
\sum_{\substack{a_2'\Lambda_2'\\(\lambda_2')}}
\cg{\Lambda_2'}{\lambda_2'}{\Lambda_T}{\lambda_T}{\Lambda_2}{\lambda_2}
\rme[2]{\Gamma_2\\a_2\Lambda_2}{T^{\Lambda_T}}{\Gamma_2\\a_2'\Lambda_2'}
\cg{\Lambda_1}{\lambda_1}{\Lambda_2'}{\lambda_2'}{\Lambda}{\lambda}
\cg{\Gamma_1}{a_1\Lambda_1}{\Gamma_2}{a_2'\Lambda_2'}{\rho\Gamma}{a\Lambda}.
\end{multline}
It now remains to eliminate the coupling coefficients of $H$, and thus
all reference to weights.  The orthogonality relations allow these
coefficients to be moved to the right hand side, resulting in sums of
quadruple products of coupling coefficients.  These sums are
recognized as recoupling coefficients of $H$, specifically, the
``unitary $6$-$\Lambda$ symbols'', or transformation brackets between
basis states in the coupling schemes
$[(\Lambda_1\Lambda_2)^{\Lambda_{12}}\Lambda_3]^{\Lambda}$ and
$[\Lambda_1(\Lambda_2\Lambda_3)^{\Lambda_{23}}]^{\Lambda}$, which are
given by
\begin{equation}
\label{eqn-usixj}
\usixj{\Lambda_1}{\Lambda_2}{\Lambda_{12}}{\Lambda_3}{\Lambda}{\Lambda_{23}}
=
\sum_{\substack{\lambda_1\lambda_2\lambda_3\\\lambda_{12}\lambda_{13}}}
\cg{\Lambda_1}{\lambda_1}{\Lambda_2}{\lambda_2}{\Lambda_{12}}{\lambda_{12}}
\cg{\Lambda_{12}}{\lambda_{12}}{\Lambda_3}{\lambda_3}{\Lambda}{\lambda}
\cg{\Lambda_2}{\lambda_2}{\Lambda_3}{\lambda_3}{\Lambda_{23}}{\lambda_{23}}
\cg{\Lambda_1}{\lambda_1}{\Lambda_{23}}{\lambda_{23}}{\Lambda}{\lambda}.
\end{equation}
The $6$-$\Lambda$ symbol is nonvanishing only if the Clebsch-Gordan
series relations $\Lambda_1\otimes\Lambda_2\rightarrow\Lambda_{12}$,
$\Lambda_{12}\otimes\Lambda_3\rightarrow\Lambda$,
$\Lambda_2\otimes\Lambda_3\rightarrow\Lambda_{23}$, and
$\Lambda_1\otimes\Lambda_{23}\rightarrow\Lambda$ are satisfied.  Let
$\Phi(\Lambda_1\Lambda_2;\Lambda)$ denote the phase factor incurred
by interchange of the first and second irreps in a coupling
coefficient of $H$, {\it i.e.},
$\smallcg{\Lambda_2}{\lambda_2}{\Lambda_1}{\lambda_1}{\Lambda}{\lambda}=
\Phi(\Lambda_2\Lambda_1;\Lambda)\smallcg{\Lambda_1}{\lambda_1}{\Lambda_2}{\lambda_2}{\Lambda}{\lambda}$.
Then the condition~(\ref{eqn-racah-semireduced}) becomes, with 
labels renamed for simplicity,
\begin{multline}
\label{eqn-racah-reduced}
\sum_{a}
\rme[2]{\Gamma\\a\Lambda}{T^{\Lambda_T}}{\Gamma\\a'\Lambda'}
\cg{\Gamma_1}{a_1\Lambda_1}{\Gamma_2}{a_2\Lambda_2}{\rho\Gamma}{a\Lambda}
\\=
\sum_{a_1'\Lambda_1'}\Phi(\Lambda_1\Lambda_2;\Lambda)\Phi(\Lambda_1'\Lambda_2;\Lambda')
\usixj{\Lambda_2}{\Lambda_1'}{\Lambda'}{\Lambda_T}{\Lambda}{\Lambda_1}
\rme[2]{\Gamma_1\\a_1\Lambda_1}{T^{\Lambda_T}}{\Gamma_1\\a_1'\Lambda_1'}
\cg{\Gamma_1}{a_1'\Lambda_1'}{\Gamma_2}{a_2\Lambda_2}{\rho\Gamma}{a'\Lambda'}
\\+
\sum_{a_2'\Lambda_2'}
\usixj{\Lambda_1}{\Lambda_2'}{\Lambda'}{\Lambda_T}{\Lambda}{\Lambda_2}
\rme[2]{\Gamma_2\\a_2\Lambda_2}{T^{\Lambda_T}}{\Gamma_2\\a_2'\Lambda_2'}
\cg{\Gamma_1}{a_1\Lambda_1}{\Gamma_2}{a_2'\Lambda_2'}{\rho\Gamma}{a'\Lambda'},
\end{multline}
expressed entirely in terms of the reduced coupling coefficients to be
calculated, reduced matrix elements, and recoupling coefficients of
the lower algebra $H$.

For the important special case in which $H$ is the angular momentum
algebra $\grpso{3}\sim\grpsu{2}$, the
relation~(\ref{eqn-racah-reduced}) becomes
\begin{multline}
\label{eqn-racah-reduced-so3}
\sum_{a}
\rme[2]{\Gamma\\aJ}{T^{(J_T)}}{\Gamma\\a'J'}_{\grpso{3}}
\cg{\Gamma_1}{a_1J_1}{\Gamma_2}{a_2J_2}{\rho\Gamma}{aJ}
\\=
\sum_{a_1'J_1'}
(-)^{J_2+J_T+J'}\Jhat \Jhat'
\sixj{J_2}{J_1'}{J'}{J_T}{J}{J_1}
\rme[2]{\Gamma_1\\a_1J_1}{T^{(J_T)}}{\Gamma_1\\a_1'J_1'}_{\grpso{3}}
\cg{\Gamma_1}{a_1'J_1'}{\Gamma_2}{a_2J_2}{\rho\Gamma}{a'J'}
\\+
\sum_{a_2'J_2'}
(-)^{J_1+J_2'+J} \Jhat \Jhat'
\sixj{J_1}{J_2'}{J'}{J_T}{J}{J_2}
\rme[2]{\Gamma_2\\a_2J_2}{T^{(J_T)}}{\Gamma_2\\a_2'J_2'}_{\grpso{3}}
\cg{\Gamma_1}{a_1J_1}{\Gamma_2}{a_2'J_2'}{\rho\Gamma}{a'J'},
\end{multline}
where $\Jhat\equiv(2J+1)^{1/2}$.  Note that the customary
form\cite{edmonds1960:am} of the Wigner-Eckhart theorem for
$\grpso{3}$ is defined in terms of a $3$-$J$ symbol rather than a
Clebsch-Gordan coefficient.  This results in a reduced matrix element
which differs in normalization and phase from the definition implied
by the generic statement of the Wigner-Eckhart theorem
in~(\ref{eqn-we}).  The reduced matrix elements under these two
conventions are related by
$\rme{J_3}{T^{(J_2)}}{J_1}_{\grpso{3}}=(-)^{2J_2}\Jhat_3
\rme{J_3}{T^{(J_2)}}{J_1}$.

\subsection{Solution of the homogeneous system}
\label{sec-solution}

The condition~(\ref{eqn-racah-reduced}) yields a different relation
among specific reduced coupling coefficients for each choice of values
for the four irrep labels $a_1\Lambda_1$, $a_2\Lambda_2$, $\Lambda$,
and $a'\Lambda'$ (the multiplicity index $a$ is summed over).  If there
are $N$ coupling coefficients for the coupling
$\Gamma_1\otimes\Gamma_2\rightarrow\rho\Gamma$, then the relations
obtained from~(\ref{eqn-racah-reduced}) constitute a linear,
homogeneous system of equations in $N$ unknowns for these coupling
coefficients.

Note that the most familiar and traditional approach to extracting
coupling coefficients, after obtaining some set of relations among
them, is to proceed by recurrence (\textit{e.g.}, in the familiar case
of $\grpso{3}$\cite{edmonds1960:am} and in the ``building-up
process'',\cite{wybourne1974:groups} as well as in prior applications of
Racah's method to higher
algebras\cite{hecht1965:so5-wigner,han1993:so5-coupling}).  That is, a
seed value is given for one coupling coefficient, and further
coefficients are deduced inductively (one by one) from those already
obtained.
\begin{figure}
\begin{center}
\includegraphics*[width=0.9\hsize]{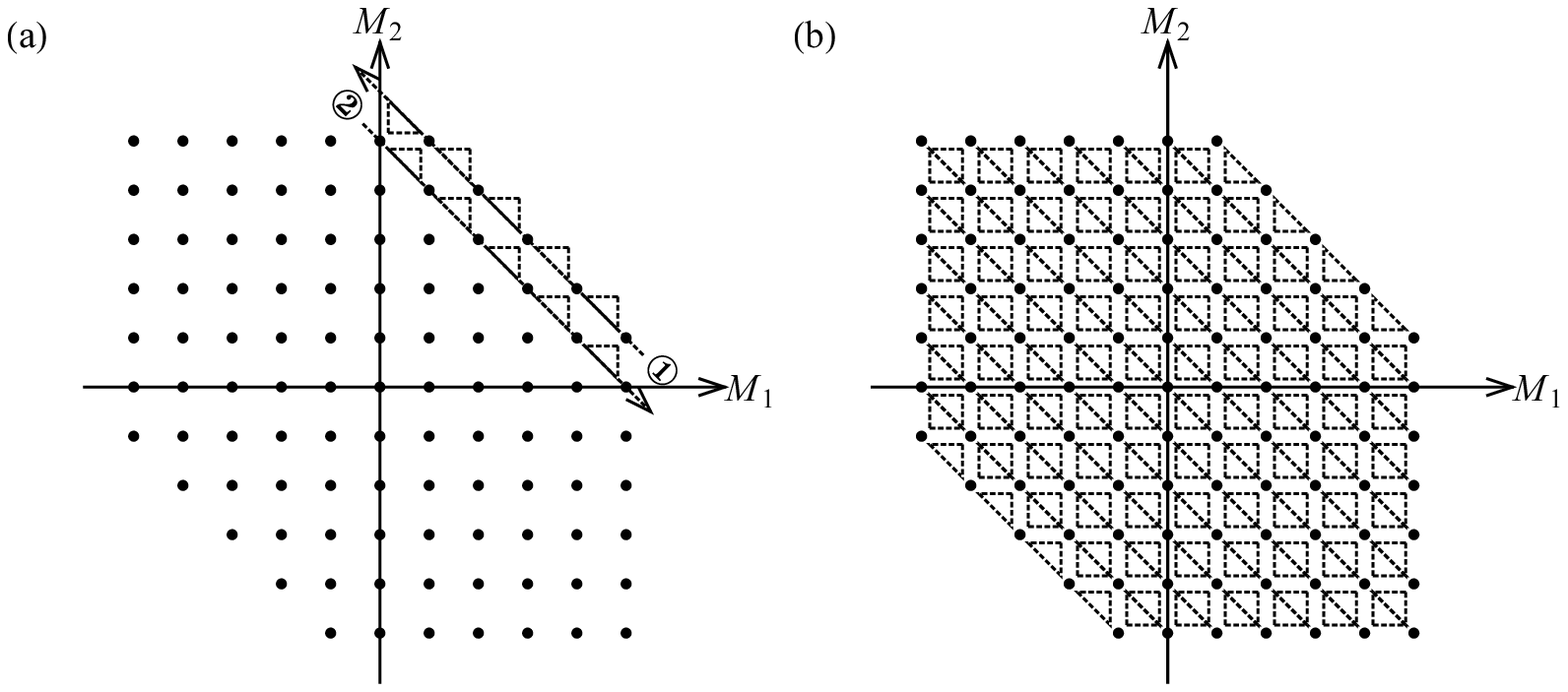}
\end{center}
\caption{The classic problem of constructing the $\grpso{3}$ Clebsch-Gordan
coefficients $\smallcg{J_1}{M_1}{J_2}{M_2}{J}{M_1+M_2}$ by use of the
relations~(\ref{eqn-jpm-relations}).  Dots indicate allowed non-zero
coefficients.  Coefficients at the vertices of a dashed triangle are
connected by~(\ref{eqn-jpm-relations}).  (a)~The conventional
recurrence approach, in which coefficients are calculated inductively
from a seed coefficient, making use of relations which in some cases
invoke known-zero Clebsch-Gordan coefficients. (b)~A full set of
relations among allowed Clebsch-Gordan coefficients, yielding a
linear, homogeneous system of equations in the Clebsch-Gordan
coefficients.  
}
\label{fig-so3-recurrence}
\end{figure}

A recurrence approach is indeed natural in the case of $\grpso{3}$.
The relations obtained by considering the actions of $J_\pm$ are
\begin{equation}
\label{eqn-jpm-relations}
K_{\pm}(JM)\smallcg{J_1}{M_1}{J_2}{M_2}{J}{M\pm1} =
K_{\pm}(J_1M_1\mp1)\smallcg{J_1}{M_1\mp1}{J_2}{M_2}{J}{M} +
K_{\pm}(J_2M_2\mp1)\smallcg{J_1}{M_1}{J_2}{M_2\mp1}{J}{M},
\end{equation}
where $M=M_1+M_2\mp1$, and
$K_{\pm}(JM)\equiv\me{JM\pm1}{J_\pm}{JM}=[(J\mp M)(J\pm M+1)]^{1/2}$
is the generator matrix element.  These relations connect at most
three coupling coefficients, and a natural order for traversing the
coefficients can easily be chosen, such that only one unknown arises
at each step, as illustrated in Fig.~\ref{fig-so3-recurrence}(a).
[This is accomplished by involving certain known-zero or ``forbidden''
Clebsch-Gordan coefficients, represented by the triangle vertices
without dots in Fig.~\ref{fig-so3-recurrence}(a), in the relations.]
Since classic treatments (\textit{e.g.}, Ref.~\onlinecite{sakurai1994:qm})
apply orthonormality relations interspersed with the recurrence
relations at intermediate stages of the calculation, we stress that
recourse to orthonormality conditions is not actually necessary.  It
may be seen from the figure that all coefficients are accessible by
the relations~(\ref{eqn-jpm-relations}).  In anticipation of the
treatment of higher algebras, we also observe that all allowed
coefficients may be connected by the relations directly, without
involving any forbidden coefficients, as in
Fig.~\ref{fig-so3-recurrence}(b).  This yields a system of equations
which fully determines the coefficients, to within an overall phase
and normalization, although in this case the system is not amenable
to solution by recursive calculation of successive coefficients from a
single seed coefficient.

For higher algebras, many irreps of $H$ may be connected by the
generator $T^{(\Lambda_T)}$, and therefore each relation obtained
from~(\ref{eqn-racah-reduced}) may involve many unknown coupling
coefficients.  A simple recurrence pattern, as in
Fig.~\ref{fig-so3-recurrence}(a), may be impractical to devise.  A
more generally applicable and straightforward approach is to directly
solve the linear, homogeneous system of equations for the unknown
coupling coefficients, by standard linear algebraic methods, {\it e.g.}, 
Euler row reduction.\cite{fn-nullmethod}

Let us therefore summarize the system of equations which must be
constructed and solved.  The $N$ unknown coupling coefficients for
$\Gamma_1\otimes\Gamma_2\rightarrow\rho\Gamma$ may be labeled with a
single counting index as
\begin{math}
C_i\equiv\smallcg{\Gamma_1}{a_1\Lambda_1}{\Gamma_2}{a_2\Lambda_2}{\rho\Gamma}{a\Lambda}
\end{math}
with $i=1,\ldots,N$.  Each value of the index $i$ therefore designates
a specific combination $(a_1\Lambda_1a_2\Lambda_2a\Lambda)$.  The
numerical coefficients of the unknown quantities $C_i$ in the
relations~(\ref{eqn-racah-reduced}) do {\it not} depend upon the outer
multiplicity index $\rho$, to be discussed further below.  For each
generator $T^{(\Lambda_T)}$ in $G$ but not in $H$, and for each
quadruplet of irrep labels $(a_1\Lambda_1a_2\Lambda_2\Lambda a'\Lambda')$,
the relation~(\ref{eqn-racah-reduced}) yields an equation (which we
label by a counting index $k$) of the form $\sum_{i=1}^N a_{ki}C_i=0$,
\textit{i.e.}, linear and homogeneous in the $C_i$.  The equation is
nonnull (\textit{i.e.}, some of the coefficients $a_{ki}$ are
nonvanishing) only if the Clebsch-Gordan series conditions
$\Lambda_1\otimes\Lambda_2\rightarrow\Lambda$ and
$\Lambda'\otimes\Lambda_T\rightarrow\Lambda$ are
met.\cite{fn-terms}

  The resulting equations must be aggregated to yield the full system,
  which may be expressed in matrix form as
\begin{equation}
\label{eqn-system-matrix}
\begin{matrix}
& {
\text{Coupling coefficient}\atop
\xrightarrow{(a_1\Lambda_1a_2\Lambda_2a\Lambda)}
}
\\
\rotatebox[origin=c]{90}{
\begin{math}
{
\text{Racah relation}
\atop
\xleftarrow{(a_1\Lambda_1a_2\Lambda_2\Lambda a' \Lambda')}
}
\end{math}
} &
\underbrace{
\begin{bmatrix}
&\vdots &\\
\cdots & a_{ki}&\cdots\\
&\vdots &\\
\end{bmatrix}
}_{\equiv A} &
\begin{bmatrix}
C_1\\\vdots\\C_N
\end{bmatrix}
=
\begin{bmatrix}
0\\
\vdots\\
0
\end{bmatrix}
\end{matrix}
.
\end{equation}
Normally, it suffices to consider the conditions obtained with
$(\Lambda_1\Lambda_2\Lambda \Lambda')$ such that
$\Gamma_1\rightarrow\Lambda_1$, $\Gamma_2\rightarrow\Lambda_2$,
$\Gamma\rightarrow\Lambda$, and $\Gamma\rightarrow\Lambda'$, that is,
relations involving only ``allowed'' coupling coefficients, as in
Fig.~\ref{fig-so3-recurrence}(b).  However, in certain exceptional
cases,\cite{fn-identitycoupling} additional conditions involving
known-zero coupling coefficients may be necessary, analogous to
Fig.~\ref{fig-so3-recurrence}(a). These may be obtained,
\textit{e.g.}, by considering some $\Lambda'$ with 
$\Gamma\nrightarrow\Lambda'$.

The problem of solving this linear homogeneous system of
equations~(\ref{eqn-system-matrix}) is equivalent to finding the null
vector (or vectors) of the matrix $A$ appearing on the left hand side
of~(\ref{eqn-system-matrix}).  In general, there may be many more rows
(equations) than columns (unknown coupling coefficients).  However,
these rows are not linearly independent.  In the case where
$\Gamma_1\otimes\Gamma_2\rightarrow\Gamma$ is free of outer
multiplicity, the matrix $A$ can be expected to be of rank $N-1$.  The
null vector is then uniquely determined, to within normalization and
phase, and its entries are the coupling coefficients
$[C_1\,C_2\,\cdots\,C_N]$.  The proper normalization, yielding
coefficients satisfying the condition~(\ref{eqn-ortho-bra-sum}), is
obtained by evaluating
\begin{equation}
\label{eqn-ortho-N}
\scrN^2\equiv\sum_{\substack{i\\(\text{same~}a\Lambda)}} C_i^2.
\end{equation}
That is, the summation runs over the subset of entries $C_i$ sharing
the same value for $a\Lambda$.  An identical result for $\scrN$ must
be obtained, regardless of the choice of $a\Lambda$, provided
$\Gamma\rightarrow a\Lambda$.  (In fact, the requirement of equality
may be used as an internal consistency check on the calculation.)  The
normalized coupling coefficients are then obtained by dividing the
null vector by $\scrN$.

An overall sign remains to be chosen for the entire set of coupling
coefficients for $\Gamma_1\otimes\Gamma_2\rightarrow\Gamma$.  For
instance, Refs.~\onlinecite{hecht1965:so5-wigner,hemenger1970:so5-quasispin}
suggest a``generalized Condon-Shortley phase
convention'',
such that $\smallcg{\Gamma_1}{\Lambda_{1m}}{\Gamma_2}{a_2\tilde{\Lambda}_{2}}{\Gamma}{\Lambda_{m}}>0$,
that is, a positive value is adopted for the coupling coefficient
involving the highest weight irreps of $H$ contained in $\Gamma_1$ and
$\Gamma$ and the highest weight irrep $\tilde{\Lambda}_{2}$ consistent
with these.

More generally, when the coupling
$\Gamma_1\otimes\Gamma_2\rightarrow\Gamma$ has outer multiplicity $D$
($\rho=1,\ldots,D$), the matrix $A$ may be expected to be of rank
$N-D$.  That is, the system of equations given by~(\ref{eqn-system-matrix})
yields $D$ linearly independent null vectors (or its null space has
dimension $D$).  The null vectors obtained by Euler row reduction must
be orthonormalized,\cite{fn-inner}
\textit{e.g.}, by the Gram-Schmidt procedure, to yield a set of
coupling coefficients satisfying the orthonormality
relation~(\ref{eqn-ortho-bra-sum}).  Note that the appropriate inner
product for this orthonormalization is {\it not} the standard vector
dot product on $\bbR^N$.  Rather, if we label the entries of each null
vector $\vec{C}_{\rho}$ as $[C_{\rho1}\,C_{\rho2}\,\cdots\,C_{\rho N}]$,
then the inner product to be used for orthonormalization is
\begin{equation}
\label{eqn-ortho-M}
\scrM_{\rho'\rho}\equiv\sum_{\substack{i\\(\text{same~}a\Lambda)}} C_{\rho' i}C_{\rho i}.
\end{equation}
The same value of $\scrM_{\rho'\rho}$ is obtained regardless of the
choice of $a\Lambda$ used in evaluating the sum.  (Again, requirement
of this equality provides an internal consistency check on the
calculation.)  The orthonormal coupling coefficients are then simply
the entries of the orthonormalized null vectors.

When an outer multiplicity is present, it should be noted that the
coupling coefficients are defined only to within a unitary
transformation, arising from the arbitrariness inherent in defining
the resolution of the outer multiplicity, \textit{i.e.}, in choosing
the basis states $\ket{\rho\Gamma\cdots}$ ($\rho=1,\ldots,D$) spanning
the $D$-dimensional space of irreps of type $\Gamma$.  In the present
calculational procedure, the freedom in resolution of the multiplicity
is manifested in the freedom to choose different sets of orthogonal
basis vectors for the null space of $A$.

\section{\boldmath Coupling coefficients for $\grpso{5}$ in the
canonical basis}
\label{sec-so5}

\subsection{Overview}
\label{example-overview}

For a concrete example of the application of Racah's method in terms
of reduced coupling coefficients, as developed in
Sec.~\ref{sec-method}, we consider the calculation of coupling
coefficients for $\grpso{5}$, reduced with respect to the canonical
subalgebra $\grpso{4}$.  That is, we have $\grpso{5}\supset\grpso{4}$
as the algebras $G\supset H$.  Both essential criteria for application
of the method are met: (1)~the coupling and recoupling coefficients
(Wigner calculus) for $\grpso{4}$ are
known,\cite{biedenharn1961:so4-wigner} and (2)~the reduced matrix
elements of the $\grpso{5}$ generators, considered as tensor operators
with respect to $\grpso{4}$, are also known.\cite{hecht1965:so5-wigner,kemmer1968:so5-irreps-1}

The algebra $\grpso{5}$ contains several subalgebra chains, involving
distinct $\grpso{3}\sim\grpsu{2}$ subalgebras,
\begin{equation}
\label{eqn-so5-chains}
\begin{aligned}
\subqn{\grpso{5}}{[l_1l_2]} 
&\supset\subqn{\grpso{4}}{[pq]}\supset\subqn{\grpso[J]{3}}{J}\supset\subqn{\grpso[J]{2}}{M_J}
&\quad&(\chain{I})
\\
&\supset\subqn{\grpso{4}}{}\sim\subqn{\grpso[X]{3}}{X}\otimes\subqn{\grpso[Y]{3}}{Y}\supset\subqn{\grpso[X]{2}}{M_X}\otimes\subqn{\grpso[Y]{2}}{M_Y}
&&(\chain{I'})
\\
&\subqnsupset{\kappa}\subqn{\grpu[N]{1}}{M_S}\otimes\subqn{\grpso[T]{3}}{T}\supset\subqn{\grpso[T]{2}}{M_T}
&&(\chain{II})
\\
&\subqnsupset{\alpha}\subqn{\grpso[L]{3}}{L}\supset\subqn{\grpso[L]{2}}{M_L}
&&(\chain{III}),
\end{aligned}
\end{equation}
where the irrep label has been noted beneath each subalgebra.
Branching multiplicity labels are indicated by $\kappa$ and $\alpha$
in the last two chains.  Chain~(\chain{I}) is the standard canonical
chain, while in~(\chain{I'}) the canonical $\grpso{4}$ subalgebra is
reexpressed using the isomorphism
$\grpso{4}\sim\grpso{3}\otimes\grpso{3}$.  [As far as definition of
reduced coupling coefficients is concerned, the two chains~(\chain{I})
and~(\chain{I'}) are equivalent, but branching rules, coupling
coefficients, \textit{etc.}, are simpler when expressed with respect
to the latter chain~(\chain{I'}).]  The prerequisite definitions and algebraic
results are summarized in Sec.~\ref{sec-alg}, and the calculation of
reduced coupling coefficients for the canonical chain is discussed in
Sec.~\ref{sec-canonical}. 

Physical applications require the coupling coefficients of $\grpso{5}$
reduced with respect to the noncanonical subalgebras of
chains~(\chain{II}) and~(\chain{III}).  The isospin algebra
$\grpso[T]{3}$ of chain~(\chain{II}) is the relevant subalgebra for
the description of proton-neutron
pairing.\cite{helmers1961:shell-sp,flowers1964:quasispin,ichimura1965:seniority-isospin,ginocchio1965:so5-quasispin,hecht1967:so5-shell-wigner,engel1996:so5-isovector-pairing,sviratcheva2005:isospin-breaking-sp4,sviratcheva2006:realistic-symmetries}
In this context, the $\grpso{5}$ generators arise as quasispin
operators for pairing of protons and neutrons occupying the same
$j$-shell.  On the other hand, the ``physical'' or ``geometric''
angular momentum subalgebra $\grpso[L]{3}$ of chain~(\chain{III}) is
the relevant subalgebra for application to systems of spin-$2$
bosons\cite{arima1976:ibm-u5,iachello1987:ibm} or the nuclear
collective
model.\cite{bohr1998:v2,eisenberg1987:v1,rowe2004:spherical-harmonics,rowe2009:acm}
Explicit
constructions of these subalgebras and further algebraic properties
for the noncanonical chains are detailed in
Appendix~\ref{app-transform}, where the transformation between
canonical and noncanonical bases is considered.

\subsection{Definitions and algebraic properties}
\label{sec-alg}

Let us begin with a concise but comprehensive summary of the
construction of $\grpso{5}$ and the algebraic properties
needed for the application of Racah's method.  Such a review is
particularly necessary since notations and conventions for nearly all
aspects of the treatment of $\grpso{5}$ vary widely (\textit{e.g.},
Refs.~\onlinecite{hecht1965:so5-wigner,kemmer1968:so5-irreps-1,behrends1962:groups-strong,wybourne1974:groups}),
and phases and normalizations play an essential role in the
calculation of coupling coefficients.

The basic construction proceeds from the generators of rotation, 
\begin{equation}
\label{eqn-L-defn}
L_{rs}\equiv -i(x_r\partial_s-x_s\partial_r).
\end{equation}
These operators are Hermitian ($L_{rs}^\dagger=L_{rs}$), are antisymmetric
in the indices, and have commutators
\begin{equation}
\label{eqn-L-comm}
[L_{pq},L_{rs}]= -i(\delta_{qr}L_{ps}+\delta_{ps}L_{qr}+\delta_{sq}L_{rp}+\delta_{rp}L_{sq}).
\end{equation}
First, for $\grpso{4}$, let $J_r\equiv\half\varepsilon_{rst}L_{st}$ and
$N_r\equiv L_{r4}$ ($1\leq r,s,t\leq3$), \textit{i.e.},
\begin{equation}
\label{eqn-so4-gen}
\begin{aligned}
J_1&=L_{23}&J_2&=L_{31}&J_3&=L_{12}\\
N_1&=L_{14}&N_2&=L_{24}&N_3&=L_{34}.
\end{aligned}
\end{equation}
Then the $J_r$ span the usual three-dimensional angular
momentum algebra, which we denote by $\grpso[J]{3}$.  The $J_r$ and $N_r$
together span $\grpso{4}$, with commutators
$[J_r,J_s]=i\varepsilon_{rst}J_t$, $[N_r,N_s]=i\varepsilon_{rst}J_t$,
and $[J_r,N_s]=i\varepsilon_{rst}N_t$.  The standard Cartan weight
operators for $\grpso{4}$ are $J_3$ and $N_3$.  An $\grpso{4}$ irrep
is labeled by the highest weight defined by these operators, which is of the form
$[pq]$, with $p\geq\abs{q}$, both integer or both odd half integer.

The isomorphism $\grpso{4}\sim\grpso[X]{3}\otimes\grpso[Y]{3}$ is
realized by taking
\begin{equation}
\label{eqn-so4-iso}
X_k\equiv\half(J_k+N_k) \quad Y_k\equiv\half(J_k-N_k),
\end{equation}
so $[X_r,X_s]=i\varepsilon_{rst}X_t$, $[Y_r,Y_s]=i\varepsilon_{rst}Y_t$,
and $[X_r,Y_s]=0$.  The ladder operators for each $\grpso{3}$ algebra
are thus $X_\pm\equiv X_1\pm i X_2$ and $Y_\pm\equiv Y_1\pm i Y_2$.
The natural Cartan weight operators in this scheme are then the
$\grpso{3}$ angular momentum projections $X_0\equiv X_3$
and $Y_0\equiv Y_3$, defining weight labels $M_X$ and $M_Y$.  An
$\grpso{4}$ irrep is then labeled by the highest weight $(XY)$,
\textit{i.e.}, the angular momenta associated with the $\grpso[X]{3}$
and $\grpso[Y]{3}$ subalgebras.  The $\grpso[X]{3}\otimes\grpso[Y]{3}$
irrep labels are related to the standard $\grpso{4}$ labels by
$X=\half(p+q)$ and $Y=\half(p-q)$ or, conversely, $[p,q]=[X+Y,X-Y]$.
Note that the canonical $\grpso[J]{3}$ is obtained as the sum angular
momentum algebra of $\grpso[X]{3}$ and $\grpso[Y]{3}$, since
$J_k=X_k+Y_k$, and thus the basis states reducing chains~(\chain{I})
and~(\chain{I'}) are related to each other by ordinary angular
momentum coupling.

The Clebsch-Gordan series and coupling and recoupling coefficients for
$\grpso{4}$ follow immediately from the $\grpso{3}\otimes\grpso{3}$
structure,\cite{biedenharn1961:so4-wigner} most transparently with
the $(XY)$ labeling scheme for the irreps.  The weights contained
within $(XY)$ are $M_X=-X,\ldots,X-1,X$ and $M_Y=-Y,\ldots,Y-1,Y$.
The Clebsch-Gordan series is given by application of the triangle
inequality separately to each of the $\grpso{3}$ algebras, that is, for $(X_1Y_1)\otimes(X_2Y_2)\rightarrow(XY)$,
$X=\abs{X_1-X_2},\abs{X_1-X_2}+1,\ldots,X_1+X_2$ and $Y=\abs{Y_1-Y_2},\abs{Y_1-Y_2}+1,\ldots,Y_1+Y_2$.
Hence, no inner or outer multiplicities are obtained for $\grpso{4}$.
Coupling coefficients factorize into products of ordinary $\grpso{3}$
Clebsch-Gordan coefficients, as
\begin{equation}
\label{eqn-so4-coupling}
\cg{(X_1Y_1)}{M_{X1}M_{Y1}}{(X_2Y_2)}{M_{X2}M_{Y2}}{(XY)}{M_{X}M_{Y}}
=
\cg{X_1}{M_{X1}}{X_2}{M_{X2}}{X}{M_{X}}
\cg{Y_1}{M_{Y1}}{Y_2}{M_{Y2}}{Y}{M_{Y}}.
\end{equation}
By inspection of~(\ref{eqn-usixj}), it is immediately apparent that
the recoupling coefficients factorize as well, as
\begin{multline}
\label{eqn-so4-recoupling}
\usixj{(X_1Y_1)}{(X_2Y_2)}{(X_{12}Y_{12})}{(X_3Y_3)}{(XY)}{(X_{23}Y_{23})}
=
\usixj{X_1}{X_2}{X_{12}}{X_3}{X}{X_{23}}
\usixj{Y_1}{Y_2}{Y_{12}}{Y_3}{Y}{Y_{23}}\\
=
(-)^{X_1+X_2+X_3+X}
(-)^{Y_1+Y_2+Y_3+Y}
\hat{X}_{12}\hat{X}_{23}
\hat{Y}_{12}\hat{Y}_{23}
\sixj{X_1}{X_2}{X_{12}}{X_3}{X}{X_{23}}
\sixj{Y_1}{Y_2}{Y_{12}}{Y_3}{Y}{Y_{23}}.
\end{multline}
An equivalent result is given with standard $\grpso{4}$ labels in
Ref.~\onlinecite{iachello1991:tetratomic}.  However, note that the result is considerably more
cumbersome to derive if one uses the standard canonical
chain~(\chain{I}).\cite{ding2003:so4-recoupling-tetratomic}

The algebra $\grpso{5}$ includes the additional four generators
$L_{r5}$ ($r=1,\ldots,4$).  A tensor operator with respect to
$\grpso{4}$ is simply a simultaneous spherical tensor with respect to
both the $\grpso[X]{3}$ and $\grpso[Y]{3}$ algebras, \textit{i.e.}, a
spherical ``bitensor''.  For the $\grpso{5}$ generators, we have
bitensor expressions\cite{fn-so5gen}
\begin{equation}
\label{eqn-so5-gen}
\begin{aligned}
X^{(10)}_{\pm10}&\equiv
X_{\pm1}=\mp\tfrac1{2\sqrt{2}}[(L_{23}+L_{14})\pm i(L_{31}+L_{24})]
&
X^{(10)}_{00}&\equiv X_{0}=\half(L_{12}+L_{34})
\\
Y^{(01)}_{0\pm1}&\equiv Y_{\pm1}=\mp\tfrac1{2\sqrt{2}}[(L_{23}-L_{14})\pm i(L_{31}-L_{24})]
&
Y^{(01)}_{00}&\equiv Y_{0}=\half(L_{12}-L_{34})\\
\Tfull{+}{+}&\equiv\Tpp=-\half(L_{15}+iL_{25})
&
\Tfull{+}{-}&\equiv\Tpm=\half(L_{35}+iL_{45})\\
\Tfull{-}{+}&\equiv\Tmp=\half(L_{35}-iL_{45})
&
\Tfull{-}{-}&\equiv\Tmm=\half(L_{15}-iL_{25}).
\end{aligned}
\end{equation}
The phases are chosen so that these operators obey
$A^{(XY)\,\dagger}_{M_XM_Y}=(-)^{M_X+M_Y} A^{(XY)}_{-M_X-M_Y}$,
a generalization of the usual condition for a self-adjoint spherical
tensor.\cite{edmonds1960:am}  All commutators involving $X_\mu$ or
$Y_\mu$ have the values implied by the spherical
bitensor notation of~(\ref{eqn-so5-gen}), \textit{e.g.},
$[X_{\pm1},A^{(\lambda\lambda')}_{\mu\mu'}]=\mp[\half(\lambda\mp\mu)(\lambda\pm\mu+1)]^{1/2}A^{(\lambda\lambda')}_{(\mu\pm1)\mu'}$.
The commutators between components of $\Tss$ are given explicitly in
Table~\ref{tab-so5-comm}.
\begingroup
\begin{table}
\caption{Commutation relations between the components of $\Tss$, for the
$\grpso{5}$ generator normalization and phase conventions defined in~(\ref{eqn-so5-gen}).}
\label{tab-so5-comm}
\input{racah_tab01.tex}
\end{table}
\endgroup
\begin{figure}
\begin{center}
\includegraphics*[width=0.8\hsize]{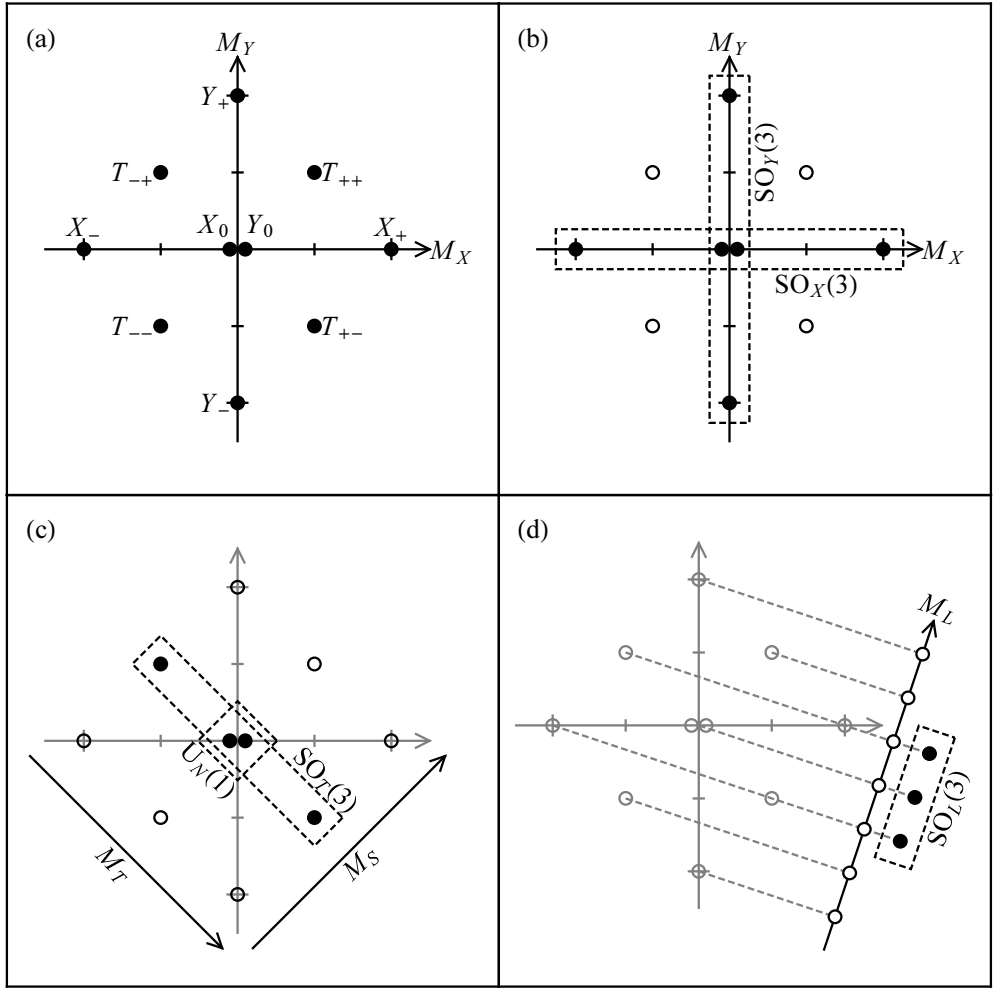}
\end{center}
\caption{
Root vector diagram for $\grpso{5}$ and its subalgebras.  (a)~The
generators of $\grpso{5}$, labeled by their Cartan weights $M_X$ and
$M_Y$.  (b)~The canonical subalgebra
$\grpso{4}\sim\grpso[X]{3}\otimes\grpso[Y]{3}$, which begins chains~(\chain{I})
and~(\chain{I'}).  (c)~The
$\grpu[N]{1}\otimes\grpso[T]{3}$ subalgebra of chain~(\chain{II}).
(d)~The physical angular momentum $\grpso[L]{3}$
subalgebra of chain~(\chain{III}).  These generators are obtained as
linear combinations of the canonical generators, as indicated by the
dashed lines, such that all generators have good
$\grpso[L]{3}\supset\grpso[L]{2}$ tensorial character.  
}
\label{fig-root-chains}
\end{figure}

The root vector diagram of $\grpso{5}$ is shown for reference in
Fig.~\ref{fig-root-chains}(a), with the generators~(\ref{eqn-so5-gen})
placed according to their $\grpso[X]{3}\otimes\grpso[Y]{3}$ weights
$(M_XM_Y)$.  The canonical subalgebra is highlighted in
Fig.~\ref{fig-root-chains}(b), and the construction of the physical
subalgebras, of 
chains~(\chain{I}) and~(\chain{III}), is indicated in
Fig.~\ref{fig-root-chains}(c,d) (see Appendix~\ref{app-transform}).

The standard Cartan highest weight labels for an $\grpso{5}$ irrep are
defined with respect to weight operators $J_3$ and $N_3$ and have the
form $[l_1l_2]$, with $l_1\geq l_2$, both integer or both odd half
integer.  It is more convenient in the present context to label
$\grpso{5}$ irreps by the highest weight defined by the
$\grpso[X]{3}\otimes\grpso[Y]{3}$ weight operators $X_0$ and $Y_0$
(following Hecht\cite{hecht1965:so5-wigner}).  The resulting label
has the form $(RS)$, with $R\geq S$, each independently either integer
or odd half integer.  This label may also be considered as
representing the angular momenta $(X_mY_m)$ of the highest weight
$\grpso[X]{3}\otimes\grpso[Y]{3}$ irrep contained in the $\grpso{5}$
irrep.  The relation to the standard labels is $R=\half(l_1+l_2)$ and
$S=\half(l_1-l_2)$.  A plethora of labeling schemes for $\grpso{5}$
irreps are in use in the physics literature,
interrelated as summarized in Table~\ref{tab-so5-label}  (even more
schemes arise if we consider
the translation to physical labels, such as reduced isospin\cite{flowers1952:jj-coupling-part1}).
\begin{table}
\caption{Labeling schemes for irreps of $\grpso{5}$ in use in the
physics literature, with relations for interconversion.}
\label{tab-so5-label}
\input{racah_tab02.tex}
\end{table}

\begin{figure}
\begin{center}
\includegraphics*[width=\hsize]{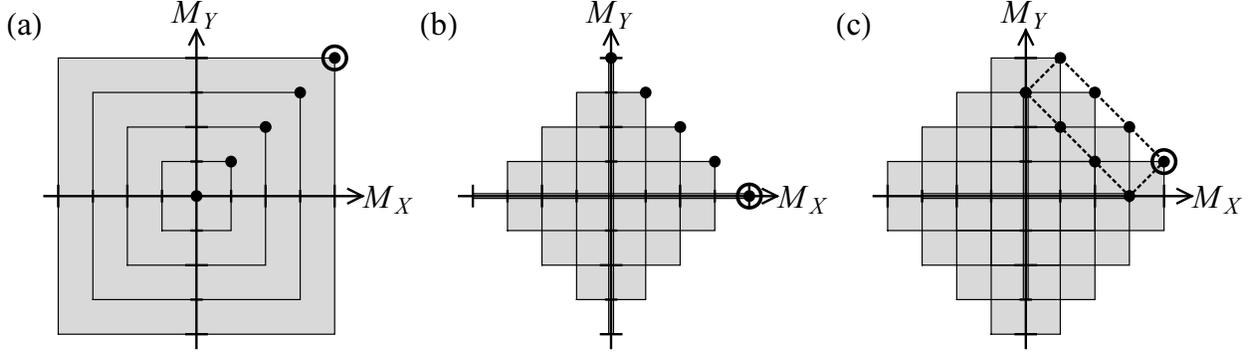}
\end{center}
\caption{
Branching diagrams for
$\grpso{5}\supset\grpso{4}$, shown for (a)~the symmetric irrep
$(22)$, (b)~the antisymmetric irrep $(20)$, and (c)~a representative
generic irrep $(\tfrac32\half)$.  The open circle indicates the
highest weight for the $\grpso{5}$ irrep, and the solid dots indicate
$\grpso{4}$ highest weights, according to branching
rule~(\ref{eqn-so5-so4-branch}).  The shaded rectangles are the
boundaries of the weight sets for these $\grpso{4}$ irreps.  The
dashed rectangle in panel~(c) is the boundary of the $\grpso{4}$
highest weight set, as discussed
in the text.}
\label{fig-weights}
\end{figure}

The branching rule for $\grpso{5}$ to $\grpso{4}$, \textit{i.e.},
$(RS)\rightarrow(XY)$, is given by\cite{hecht1965:so5-wigner,kemmer1968:so5-irreps-1}
\begin{equation}
\label{eqn-so5-so4-branch}
\begin{aligned}
X&=R-\half n-\half m\\
Y&=S+\half n-\half m,
\end{aligned}
\end{equation}
with $0\leq n\leq2(R-S)$ and $0\leq m\leq2S$, $m$ and $n$ integers.
Graphically, the $\grpso{4}$ irreps form a lattice bounded by a tilted
rectangle, as illustrated by the dotted line in
Fig.~\ref{fig-weights}(c).  The rectangle's ``right'' corner is at the
highest weight $(RS)=(X_mY_m)$, its ``bottom'' corner lies on the
$M_X$ axis, and the remaining two corners are specified by symmetry
about the line $M_X=M_Y$.  The branching rule is shown for example
irreps of $\grpso{5}$ in Fig.~\ref{fig-weights}:
symmetric [Fig.~\ref{fig-weights}(a)], antisymmetric
[Fig.~\ref{fig-weights}(b)], and generic
[Fig.~\ref{fig-weights}(c)].

The Clebsch-Gordan series for $\grpso{5}$ may be obtained by a
relatively efficient elementary approach based on the method of
weights, but ``reduced'' with respect to $\grpso{4}$.  Since the
$\grpso{5}\rightarrow\grpso{4}$ branching
rules~(\ref{eqn-so5-so4-branch}) are known, as is the $\grpso{4}$
Clebsch-Gordan series, the tabulation of
weights can be replaced by tabulation of $\grpso{4}$ irrep labels,
which then imply all the weights contained within these irreps.  To
decompose the $\grpso{5}$ outer product $(R_1S_1)\otimes(R_2S_2)$,
first the $\grpso{4}$ irreps in the branchings
$(R_1S_1)\rightarrow(X_1Y_1)$ and $(R_2S_2)\rightarrow(X_2Y_2)$ are
enumerated.  Then, for each pair of $\grpso{4}$ irreps $(X_1Y_1)$ and
$(X_2Y_2)$, the product irreps
$(X_1Y_1)\otimes(X_2Y_2)\rightarrow(XY)$ are enumerated.  The
aggregate set of these product irreps represents the $\grpso{4}$
content of $(R_1S_1)\otimes(R_2S_2)$.  The $\grpso{5}$ content can now
be extracted.  Namely, the highest weight $\grpso{4}$ label in the set
gives the highest weight $\grpso{5}$ irrep contained in
$(R_1S_1)\otimes(R_2S_2)$ [which will, incidentally, simply be the
sum of $(R_1S_1)$ and $(R_2S_2)$ as weights].
The $\grpso{4}$ content of this $\grpso{5}$ irrep can now be deleted from the
set, after which the next highest remaining $\grpso{4}$ label gives
the next highest weight $\grpso{5}$ irrep in
$(R_1S_1)\otimes(R_2S_2)$, {\it etc.}  The process is repeated until the set of
$\grpso{4}$ irreps has been exhausted.    The Clebsch-Gordan series for $\grpso{5}$
may also be obtained by group character methods (see
Refs.~\onlinecite{judd1963:operator-techniques,wybourne1974:groups}).

The remaining ingredients needed for application of Racah's method are
the $\grpso{4}$-reduced matrix elements of the ``additional''
generators of $\grpso{5}$ not contained in $\grpso{4}$, \textit{i.e.},
$\Tss$.  These were obtained in closed form, by solving certain
recurrence relations obtained from the commutators of the algebra, by
Hecht\cite{hecht1965:so5-wigner} and by Kemmer, Pursey, and Williams,\cite{kemmer1968:so5-irreps-1} as
\begin{equation}
\label{eqn-T-rme}
\begin{aligned}
\rme[2]{(RS)\\(X+\half\,Y+\half)}{\Thh}{(RS)\\(XY)}&=
\frac{
\begin{multlined}
\bigl[
(R+S-X-Y)(R+S+X+Y+3)
\\[-1ex]
\times(-R+S+X+Y+1)(R-S+X+Y+2)
\bigr]^{1/2}
\end{multlined}
}
{2\widehat{X+\half}\widehat{Y+\half}}
\\
\rme[2]{(RS)\\(X+\half\,Y-\half)}{\Thh}{(RS)\\(XY)}&=
\frac{
\begin{multlined}
\bigl[
(R+S-X+Y+1)(R+S+X-Y+2)
\\[-1ex]
\times(R-S-X+Y)(R-S+X-Y+1)
\bigr]^{1/2}
\end{multlined}
}
{2\widehat{X+\half}\widehat{Y-\half}}.
\end{aligned}
\end{equation}
These expressions are appropriate to the
Wigner-Eckart theorem normalization defined in~(\ref{eqn-we}) and the
normalization of $\Tss$ defined by~(\ref{eqn-so5-gen}).  The
remaining matrix elements, connecting $(XY)$ with
$(X-\half\,Y+\half)$ or $(X-\half\,Y-\half)$, follow from
these by
the self-adjoint property of the generators,\cite{kemmer1968:so5-irreps-1} as
\begin{equation}
\rme[2]{(RS)\\(XY)}{\Thh}{(RS)\\(X'Y')}
=
\frac{\hat{X}'\hat{Y}'}{\hat{X}\hat{Y}}
(-)^{X-X'+Y-Y'}
\rme[2]{(RS)\\(X'Y')}{\Thh}{(RS)\\(XY)}.
\end{equation}

\subsection{\boldmath Calculation of $\grpso{5}\supset\grpso{4}$ coupling coefficients}
\label{sec-canonical}

Consider calculation of the set of $\grpso{5}\supset\grpso{4}$ reduced
coupling coefficients for $(R_1S_1)\otimes(R_2S_2)\rightarrow(RS)$.
It is now straightforward to construct the terms appearing in the
Racah condition~(\ref{eqn-racah-reduced}), by use of the
chain~(\chain{I'}) branching rules, $\grpso{4}$ Clebsch-Gordan series,
$\grpso{4}$ Wigner calculus, and $\grpso{4}$-reduced matrix elements of $\Tss$,
compiled in the preceding section. For $\grpso{5}\supset\grpso{4}$,
the relation~(\ref{eqn-racah-reduced}) may be written
\begin{multline}
\label{eqn-racah-reduced-so5}
\rme[2]{(RS)\\(XY)}{\Thh}{(RS)\\(X'Y')}
\cg{(R_1S_1)}{(X_1Y_1)}{(R_2S_2)}{(X_2Y_2)}{(RS)}{(XY)}
\\
\begin{aligned}
=
&\sum_{(X_1'Y_1')}\Phi[(X_1Y_1)(X_2Y_2);(XY)]\Phi[(X_1'Y_1')(X_2Y_2);(X'Y')]
\\
&\quad
\times
\usixj{(X_2Y_2)}{(X_1'Y_1')}{(X'Y')}{(\half\half)}{(XY)}{(X_1Y_1)}
\rme[2]{(R_1S_1)\\(X_1Y_1)}{\Thh}{(R_1S_1)\\(X_1'Y_1')}
\cg{(R_1S_1)}{(X_1'Y_1')}{(R_2S_2)}{(X_2Y_2)}{(RS)}{(X'Y')}
\\
+&
\sum_{(X_2'Y_2')}
\usixj{(X_1Y_1)}{(X_2'Y_2')}{(X'Y')}{(\half\half)}{(XY)}{(X_2Y_2)}
\rme[2]{(R_2S_2)\\(X_2Y_2)}{\Thh}{(R_2S_2)\\(X_2'Y_2')}
\cg{(R_1S_1)}{(X_1Y_1)}{(R_2S_2)}{(X_2'Y_2')}{(RS)}{(X'Y')}.
\end{aligned}
\end{multline}
The system of equations for the coupling coefficients is assembled
following the approach of Sec.~\ref{sec-solution}.  A different
condition is obtained from~(\ref{eqn-racah-reduced-so5}) for each
quadruplet of $\grpso{4}$ irreps $(X_1Y_1)$, $(X_2Y_2)$, $(XY)$, and
$(X'Y')$, chosen from the branchings $(R_1S_1)\rightarrow(X_1Y_1)$,
$(R_2S_2)\rightarrow(X_2Y_2)$, $(RS)\rightarrow(XY)$, and
$(RS)\rightarrow(X'Y')$ [for couplings involving the identity irrep
$(00)$, see endnote~\onlinecite{fn-identitycoupling}].  A nonnull
condition on the coupling coefficients is obtained only if
$(X_1Y_1)\otimes(X_2Y_2)\rightarrow(XY)$ and
$(X'Y')\otimes(\half\half)\rightarrow(XY)$.  Since all quantities
involved in the conditions~(\ref{eqn-racah-reduced-so5}) are known
exactly, in the form of square roots of rational numbers, and since
Euler row reduction can be carried out in exact (\textit{i.e.},
symbolic) arithmetic, all coupling coefficients can be obtained
exactly through the present process, again as (signed) square roots of
rational numbers.

Two concrete numerical examples are provided as illustrations of the
method in Appendix~\ref{app-canonical}.  A simple low-dimensional
example is provided by the coupling 
$(\half\half)\otimes(\half0)\rightarrow(\half0)$,
which
involves only a $4\times4$ coefficient matrix, and an
example involving an outer multiplicity is provided by the coupling
$(10)\otimes(1\half)\rightarrow(1\half)$, for which the
coefficient matrix has dimensions $36\times18$.  (In the canonical
labeling scheme, these examples are
$[10]\otimes[\half\half]\rightarrow[\half\half]$ and
$[11]\otimes[\tfrac32\half]\rightarrow[\tfrac32\half]$, respectively.)

\subsection{\boldmath Noncanonical chains}
\label{sec-noncanonical}

In the previous section, it was seen how the coupling coefficients for
$\grpso{5}$ reduced with respect to the {\it canonical} chain may be
evaluated by Racah's method, as formulated in Sec.~\ref{sec-method}.
The necessary ingredients take on a particularly simple form for the
canonical chain, in that the reduced matrix elements of the generators are
given by closed form expressions~(\ref{eqn-T-rme}).  However, the
matrix elements of the generators reduced with respect to the
noncanonical isospin subalgebra [chain~(\chain{II})] and physical
angular momentum algebra [chain~(\chain{III})] are also known.
Recurrence relations for the reduced matrix elements have been
obtained from vector coherent state
realizations,\cite{hecht1982:sp4-vcs,turner2006:so5-so3-vcs} or an
elementary construction has also been
demonstrated for chain~(\chain{II}).\cite{han1993:so5-coupling}
Hence, Racah's method may be applied directly to the calculation of
chain~(\chain{II}) and chain~(\chain{III}) reduced coupling
coefficients.

Alternatively, once coupling coefficients for the canonical chain have
been obtained, the coupling coefficients for the noncanonical chains
can readily be deduced by a unitary transformation.  For instance, the
coupling coefficients of $\grpsu{3}$ reduced with respect to its
physical angular momentum subalgebra are conventionally obtained from
the canonical $\grpsu{3}\supset\grpu{1}\otimes\grpsu{2}$ coupling
coefficients\cite{draayer1973:su3-cg,rowe2000:su3-cg} through such a
process.  The transformation brackets between basis states reducing
the canonical and noncanonical chains are only known in closed form
for a few special cases.\cite{hecht1967:so5-shell-wigner}  However,
they can be obtained in a straightforward fashion, either (1)~by
diagonalizing the appropriate Casimir operator, {\it i.e.},
$\vec{T}^2$ or $\vec{L}^2$, in the canonical basis, as in
Ref.~\onlinecite{debaerdemacker2008:collective-cartan}, or (2)~by a
combination of laddering and orthogonalization operations.  The
procedure for transformation of coupling coefficients to either of the
noncanonical chains is discussed in Appendix~\ref{app-transform}.

\section{Conclusion}
\label{sec-concl}

It has been shown that Racah's method of infinitesimal generators can
be systematically generalized to the calculation of reduced coupling
coefficients for an arbitrary subalgebra chain, provided the matrix
elements of the generators (reduced with respect to the lower algebra)
and the recoupling coefficients of the lower algebra are known.  For
the algebra $\grpso{5}$, the problem of calculating coupling
coefficients for generic irreps, reduced with respect to the canonical
or noncanonical chains, is thereby completely resolved.

The specific example of $\grpso{5}$ coupling coefficients may be
considered as a prototype for the systematic calculation of coupling
coefficients for other higher algebras.  For instance, the
computational machinery for $\grpsu{3}$ is well
established\cite{draayer1973:su3-cg,rowe2000:su3-cg} and may
therefore be used as the starting point for calculation of
$\grpsp{6}\supset\grpu{3}$ reduced coupling coefficients, for the
fermion dynamical symmetry model,\cite{wu1986:fdsm} or
$\grpsp{6,\bbR}\supset\grpu{3}$ reduced coupling coefficients, for the
symplectic shell model.\cite{rosensteel1980:sp6r-shell}  The
requisite generator matrix elements for $\grpsp{6}$ and
$\grpsp{6,\bbR}$ may be calculated from vector coherent state
realizations.\cite{rowe1984:sp6r-vcs,hecht1990:sp6-u3-vcs}  The
$\grpsp{6,\bbR}\supset\grpu{3}$ coupling coefficients are required,
for instance, if large-scale calculations are to be carried out in the
{\it ab initio} symplectic scheme of Dytrych {\it et
al.}\cite{dytrych2007:sp-ncsm-evidence,dytrych2008:sp-ncsm}

\begin{acknowledgments}
Discussions with C.~Bahri, T.~Dytrych, J.~P.~Draayer,
S.~De~Baerdemacker, D.~J.~Rowe, and F.~Iachello are gratefully
acknowledged. This work was supported by the US DOE (under grant
DE-FG02-95ER-40934), the US NSF (under grants NSF-PHY-0500291,
NSF-OCI-0904874, and NSF-PHY05-52843), and the Southeastern
Universities Reasearch Association (SURA).
\end{acknowledgments}

\appendix

\section{\boldmath General relations in the presence of outer
multiplicities for $H$}
\label{app-mult}

The derivation of Racah's method in terms of reduced quantities, as
given in Sec.~\ref{sec-method}, can readily be generalized to the case
in which the subalgebra $H$ has outer multiplicities, {\it i.e.}, its
Kronecker product is not simply reducible.  In this appendix, the
necessary generalizations of the algebraic relations ({\it e.g.},
Ref.~\onlinecite{wybourne1974:groups}) entering into the derivation of
Sec.~\ref{sec-method} are summarized, and the
fundamental relation~(\ref{eqn-racah-reduced}) for Racah's method is
extended to incorporate outer multiplicities of $H$.

The general form of Racah's factorization
lemma for $G\supset H$ is 
\begin{equation}
\label{eqn-factorization-mult}
\ccg{\Gamma_1}{a_1\Lambda_1}{\lambda_1}{\Gamma_2}{a_2\Lambda_2}{\lambda_2}{\rho\Gamma}{a\Lambda}{\lambda}
=
\sum_\sigma
\cg{\Lambda_1}{\lambda_1}{\Lambda_2}{\lambda_2}{\sigma\Lambda}{\lambda}
\cg{\Gamma_1}{a_1\Lambda_1}{\Gamma_2}{a_2\Lambda_2}{\rho\Gamma}{a\Lambda}_\sigma,
\end{equation}
where $\sigma$ is the multiplicity index for the coupling
$\Lambda_1\otimes\Lambda_2\rightarrow\Lambda$.
The reduced coupling coefficients satisfy
orthonormality relations
\begin{align}
\label{eqn-ortho-bra-sum-mult}
\sum_{\substack{a_1\Lambda_1a_2\Lambda_2\\\sigma}}
\cg{\Gamma_1}{a_1\Lambda_1}{\Gamma_2}{a_2\Lambda_2}{\rho\Gamma}{a\Lambda}_\sigma
\cg{\Gamma_1}{a_1\Lambda_1}{\Gamma_2}{a_2\Lambda_2}{\rho'\Gamma'}{a'\Lambda}_\sigma
&=\delta_{(\rho\Gamma)(\rho'\Gamma')}\delta_{aa'}
\intertext{and}
\label{eqn-ortho-ket-sum-mult}
\sum_{\rho\Gamma a}
\cg{\Gamma_1}{a_1\Lambda_1}{\Gamma_2}{a_2\Lambda_2}{\rho\Gamma}{a\Lambda}_{\sigma}
\cg{\Gamma_1}{a_1'\Lambda_1'}{\Gamma_2}{a_2'\Lambda_2'}{\rho\Gamma}{a\Lambda}_{\sigma'}
&=\delta_{(a_1\Lambda_1)(a_1'\Lambda_1')}\delta_{(a_2\Lambda_2)(a_2'\Lambda_2')}
\delta_{\sigma\sigma'},
\end{align}
for any irrep $\Lambda$ such that $\Gamma\rightarrow\Lambda$ and, in
the second relation, any values of the multiplicity indices $\sigma$
and $\sigma'$, which resolve $\Lambda_1\otimes\Lambda_2\rightarrow\Lambda$ and 
$\Lambda_1'\otimes\Lambda_2'\rightarrow\Lambda$, respectively.

For
the subalgebra $H$, the Wigner-Eckart theorem is of the form
\begin{equation}
\label{eqn-we-mult}
\me[3]{\Gamma'\\a'\Lambda'\\\lambda'}{T^{\Lambda_T}_{\lambda_T}}{\Gamma\\a\Lambda\\\lambda}
=
\sum_\sigma 
\cg{\Lambda}{\lambda}{\Lambda_T}{\lambda_T}{\sigma\Lambda'}{\lambda'}
\rme[2]{\Gamma'\\a'\Lambda'}{T^{\Lambda_T}}{\Gamma\\a\Lambda}_\sigma.
\end{equation}
The recoupling coefficients (unitary
$6$-$\Lambda$ symbols) of $H$ are
\begin{equation}
\label{eqn-usixj-mult}
\usixj{\Lambda_1}{\Lambda_2}{\sigma_{12}\Lambda_{12}}{\Lambda_3}{\Lambda}{\sigma_{23}\Lambda_{23}}_{\sigma\sigma'}
=
\sum_{\substack{\lambda_1\lambda_2\lambda_3\\\lambda_{12}\lambda_{13}}}
\cg{\Lambda_1}{\lambda_1}{\Lambda_2}{\lambda_2}{\sigma_{12}\Lambda_{12}}{\lambda_{12}}
\cg{\Lambda_{12}}{\lambda_{12}}{\Lambda_3}{\lambda_3}{\sigma\Lambda}{\lambda}
\cg{\Lambda_2}{\lambda_2}{\Lambda_3}{\lambda_3}{\sigma_{23}\Lambda_{23}}{\lambda_{23}}
\cg{\Lambda_1}{\lambda_1}{\Lambda_{23}}{\lambda_{23}}{\sigma'\Lambda}{\lambda}.
\end{equation}
This represents the transformation bracket between
basis states in the coupling schemes
$[(\Lambda_1\Lambda_2)^{\sigma_{12}\Lambda_{12}}\Lambda_3]^{\sigma\Lambda}$ and
$[\Lambda_1(\Lambda_2\Lambda_3)^{\sigma_{23}\Lambda_{23}}]^{\sigma'\Lambda}$.
Under interchange of the first and second irreps, the coupling
coefficients may be expected to satisfy a symmetry relation of the
form [{\it e.g.}, for $\grpsu{3}$, see Refs.~\onlinecite{draayer1973:su3-cg,escher1998:sp6r-shell-su3coupling}]
\begin{equation}
\label{eqn-cg-symm-mult}
\cg{\Lambda_2}{\lambda_2}{\Lambda_1}{\lambda_1}{\sigma\Lambda}{\lambda}
=\sum_{\sigma'}\Phi_{\sigma\sigma'}(\Lambda_2\Lambda_1;\Lambda)
\cg{\Lambda_1}{\lambda_1}{\Lambda_2}{\lambda_2}{\sigma'\Lambda}{\lambda}.
\end{equation}

Following the same arguments as in Sec.~\ref{sec-method}, the reduced
coupling coefficients for $G\supset H$ are found to satisfy a homogeneous
system of equations of the form
\begin{multline}
\label{eqn-racah-reduced-mult}
\sum_{a}
\rme[2]{\Gamma\\a\Lambda}{T^{\Lambda_T}}{\Gamma\\a'\Lambda'}_{\tau}
\cg{\Gamma_1}{a_1\Lambda_1}{\Gamma_2}{a_2\Lambda_2}{\rho\Gamma}{a\Lambda}_{\sigma}
\\=
\begin{lgathered}[t]
\sum_{\substack{a_1'\Lambda_1'\\\tau_1\sigma'\delta\delta'}}
\Phi_{\sigma\delta}(\Lambda_1\Lambda_2;\Lambda)\Phi_{\sigma'\delta'}(\Lambda_1'\Lambda_2;\Lambda')
\\\quad\times
\usixj{\Lambda_2}{\Lambda_1'}{\delta'\Lambda'}{\Lambda_T}{\Lambda}{\tau_1\Lambda_1}_{\tau\delta}
\rme[2]{\Gamma_1\\a_1\Lambda_1}{T^{\Lambda_T}}{\Gamma_1\\a_1'\Lambda_1'}_{\tau_1}
\cg{\Gamma_1}{a_1'\Lambda_1'}{\Gamma_2}{a_2\Lambda_2}{\rho\Gamma}{a'\Lambda'}_{\sigma'}
\\+
\sum_{\substack{a_2'\Lambda_2'\\\tau_2\sigma'}}
\usixj{\Lambda_1}{\Lambda_2'}{\sigma'\Lambda'}{\Lambda_T}{\Lambda}{\tau_2\Lambda_2}_{\tau\sigma}
\rme[2]{\Gamma_2\\a_2\Lambda_2}{T^{\Lambda_T}}{\Gamma_2\\a_2'\Lambda_2'}_{\tau_2}
\cg{\Gamma_1}{a_1\Lambda_1}{\Gamma_2}{a_2'\Lambda_2'}{\rho\Gamma}{a'\Lambda'}_{\sigma'}.
\end{lgathered}
\end{multline}
A different equation is obtained for each choice of
$(a_1\Lambda_1a_2\Lambda_2\Lambda a'\Lambda')$ and indices $\sigma$ and $\tau$, where
$\sigma$ resolves $\Lambda_1\otimes\Lambda_2\rightarrow\Lambda$ and
$\tau$ resolves $\Lambda'\otimes\Lambda_T\rightarrow\Lambda$.

\section{\boldmath Numerical examples for the calculation of $\grpso{5}\supset\grpso{4}$
reduced coupling coefficients}
\label{app-canonical}

For a simple numerical example of the calculation of
$\grpso{5}\supset\grpso{4}$ reduced coupling coefficients, according
to the methods of Sec.~\ref{sec-canonical}, consider the coupling
$(\half\half)\otimes(\half0)\rightarrow(\half0)$.  The coupling
coefficients to be obtained are
$\smallcg{(\shalf\,\shalf)}{(0\,0)}{(\shalf\,0)}{(0\,\shalf)}{(\shalf\,0)}{(0\,\shalf)}$,
$\smallcg{(\shalf\,\shalf)}{(\shalf\,\shalf)}{(\shalf\,0)}{(\shalf\,0)}{(\shalf\,0)}{(0\,\shalf)}$,
$\smallcg{(\shalf\,\shalf)}{(0\,0)}{(\shalf\,0)}{(\shalf\,0)}{(\shalf\,0)}{(\shalf\,0)}$,
and
$\smallcg{(\shalf\,\shalf)}{(\shalf\,\shalf)}{(\shalf\,0)}{(0\,\shalf)}{(\shalf\,0)}{(\shalf\,0)}$.
The system of four equations in four unknowns obtained
from~(\ref{eqn-racah-reduced-so5}) has the coefficient matrix shown in
Fig.~\ref{fig-system-small}.  This matrix is of rank $3$, admitting 
the null vector
\begin{displaymath}
\begin{bmatrix}
-\half & -1 & \vline&  \half & 1
\end{bmatrix},
\end{displaymath}
shown as a row vector for easier comparison with the coupling
coefficient labels across the top of Fig.~\ref{fig-system-small}.  It
remains to obtain the proper normalization (and phase, if a 
phase convention is to be enforced).   The vertical bar demarcates coupling 
coefficients involving the same product irrep $(XY)$, $(0\half)$ for
the first two coefficients and $(\half0)$ for the remaining two coefficients, that is, the
coefficients which appear together in the
normalization sum~(\ref{eqn-ortho-N}).  The squared norm, calculated
using either pair of coefficients, is $\scrN^2=\tfrac54$.  Hence, the
normalized coupling coefficients are given by
\begin{displaymath}
\begin{bmatrix}
-\sqrt{\tfrac15} & -\sqrt{\tfrac45}  &\vline& \sqrt{\tfrac15} & \sqrt{\tfrac45} 
\end{bmatrix},
\end{displaymath}
which incidentally also conforms to the generalized Condon-Shortly
phase convention.
\begin{figure}

\newlength{\sqrtwidth}
\settowidth{\sqrtwidth}{$-\sqrt{\tfrac12}$}
\newcommand{\racahtoplabel}[1]{\rotatebox{90}{\rule[-1.5ex]{0pt}{\sqrtwidth}$#1$}}
\newcommand{\racahleftlabel}[1]{#1}

\begin{center}
\input{racah_fig04.eps}
\end{center}
\caption{Coefficient matrix for the linear, homogenous system of
equations determining the
$\grpso{5}\supset\grpso{4}$ coupling coefficients for
$(\half\half)\otimes(\half0)\rightarrow(\half0)$.
}
\label{fig-system-small}
\end{figure}
\begin{turnpage} 
\begin{figure}[p]

\settowidth{\sqrtwidth}{$1$}
\newcommand{\racahtoplabel}[1]{\rotatebox{90}{\rule{0pt}{0pt}$#1$}}
\newcommand{\racahleftlabel}[1]{#1}

\begin{center}
\input{racah_fig05.eps}
\end{center}
\caption{(a)~Coefficient matrix for the linear, homogenous system of
equations determining the
$\grpso{5}\supset\grpso{4}$ coupling coefficients for
$(10)\otimes(1\half)\rightarrow(1\half)$.  The matrix has dimension
$36\times18$. Only the first few rows are
shown.  (b)~Basis vectors for the null space ($D=2$).  (c)~Orthonormal
coupling coefficient vectors, consisting of the
$\grpso{5}\supset\grpso{4}$ reduced coupling coefficients for $\rho=1$
(upper row) and $\rho=2$ (lower row).}
\label{fig-system-large}
\end{figure}
\end{turnpage}

One of the simplest cases in which an outer multiplicity occurs is in
the coupling $(10)\otimes(1\half)\rightarrow(1\half)$, which has
multiplicity $D=2$.  There are $18$
coupling coefficients for these irreps, related by a system of $36$
equations.  The first few rows of the coefficient matrix are shown in
Fig.~\ref{fig-system-large}(a).  The matrix is of rank $16$, admitting
two null vectors, given in Fig.~\ref{fig-system-large}(b).
Again, these are shown
as row vectors, and groups of coefficients sharing the same $(XY)$ are
delimited by
vertical bars.
The inner product matrix~(\ref{eqn-ortho-M}), which may be obtained using
any of these four groups of coefficients, is
\begin{displaymath}
\scrM=\begin{bmatrix}
\tfrac{15}{8} & \sqrt{\tfrac{45}{32}}\\
\sqrt{\tfrac{45}{32}} & \tfrac{31}{4}
\end{bmatrix}.
\end{displaymath}
The orthonormal null vectors
obtained by the Gram-Schmidt procedure with respect to $\scrM$ are
then given in Fig.~\ref{fig-system-large}(c). 
The entries of the upper and lower rows may be taken as
the $\grpso{5}\supset\grpso{4}$ reduced coupling coefficients for
$\rho=1$ and $2$, respectively.

\section{\boldmath Transformation of coupling coefficients}
\label{app-transform}

In this appendix, a general algorithm is outlined for determination of
the transformation brackets between canonical and
$\grpu[N]{1}\otimes\grpso[T]{3}$ [chain~(\chain{II})] bases of an
$\grpso{5}$ irrep.  These are the necessary ingredients for deducing
the chain~(\chain{II}) reduced coupling coefficients from the
canonical reduced coupling coefficients.  The analogous procedure for
transformation to the $\grpso[L]{3}$ basis [chain~(\chain{III})] is
also briefly considered.


First, let us review the relevant properties of the proton-neutron
quasispin realization of $\grpso{5}$ and the chain~(\chain{II}) basis
and branching rules.  For protons and neutrons occupying a single
level of angular momentum $j$ (degeneracy $2j+1$), consider the proton
pair quasispin operators ($X_+$, $X_0$, and $X_-$), neutron pair
quasispin operators ($Y_+$, $Y_0$, and $Y_-$), proton-neutron pair
quasispin operators ($S_+$, $S_0$, and $S_-$), and isospin operators
($T_+$, $T_0$, and $T_-$). Each set spans an angular momentum algebra,
which we denote by $\grpso[X]{3}$, $\grpso[Y]{3}$, $\grpso[S]{3}$, or
$\grpso[T]{3}$, respectively.  Explicitly,
\begin{equation}
\begin{aligned}
\label{eqn-chainII-gen}
X_+&=\tfrac12(\ad_p\cdot\ad_p) &
X_0&=-\tfrac14(\ad_p\cdot\at_p+\at_p\cdot\ad_p) &
X_-&=-\tfrac12(\at_p\cdot\at_p)\\
Y_+&=\tfrac12(\ad_n\cdot\ad_n) &
Y_0&=-\tfrac14(\ad_n\cdot\at_n+\at_n\cdot\ad_n) &
Y_-&=-\tfrac12(\at_n\cdot\at_n)\\
S_+&=\tfrac12(\ad_p\cdot\ad_n+\ad_n\cdot\ad_p) &
S_0&=-\tfrac12(\ad_p\cdot\at_p+\at_n\cdot\ad_n) &
S_-&=-\tfrac12(\at_n\cdot\at_p+\at_p\cdot\at_n)\\
T_+&=-(\ad_p\cdot\at_n) &
T_0&=-\tfrac12(\ad_p\cdot\at_p-\ad_n\cdot\at_n) &
T_-&=-(\ad_n\cdot\at_p),
\end{aligned}
\end{equation}
where $\at^{(j)}_m\equiv (-)^{j-m}a^{(j)}_{-m}$, and we define $A^{(j)}\cdot B^{(j)}
\equiv \jhat (A\times B)^{(0)}_0=\sum_m A_m \tilde{B}_m$ for
half-integer $j$.\cite{fn-sph-dot}  Thus, for instance, 
$X_+=\half\sum_m(-)^{j-m}\ad_{p,m}\ad_{p,-m}$,
$X_0=\tfrac14\sum_m(\ad_{p,m}a_{p,m}-a_{p,m}\ad_{p,m})$, and
$X_-=\half\sum_m(-)^{j-m}a_{p,-m}a_{p,m}$.
There are only ten independent
operators, since $S_0=X_0+Y_0$ and $T_0=X_0-Y_0$.  The operators
defined in~(\ref{eqn-chainII-gen}) obey the commutation relations of
the $\grpso{5}$ generators of Sec.~\ref{sec-alg}, with the
identifications
\begin{equation}
\label{eqn-sot3-reln}
S_+=-2\Tpp \quad S_-=2\Tmm \quad T_+=2\Tpm \quad T_-=2\Tmp,
\end{equation}
and thus span an $\grpso{5}$ algebra.

The generators $T_+$, $T_0$, $T_-$, and $S_0$ span the
$\grpu[N]{1}\otimes\grpso[T]{3}$ subalgebra of $\grpso{5}$, as shown
in Fig.~\ref{fig-root-chains}(c).  Here $\grpu[N]{1}$
is the one-dimensional algebra of $S_0$, so denoted in recognition
of the relation of this operator to the total proton-neutron number
operator $N=N_p+N_n$.
The natural weights for chain~(\chain{II}), $M_S$ and $M_T$, are
related to the chain~(\chain{I'}) weights by $M_S=M_X+M_Y$ and
$M_T=M_X-M_Y$.\cite{fn-weights-I-II}  For the proton-neutron
quasispin realization of $\grpso{5}$ defined
in~(\ref{eqn-chainII-gen}), the weights are simply related to proton
and neutron occupation numbers by $M_X=\half(N_p-\Omega)$ and
$M_Y=\half(N_n-\Omega)$, or $M_S=\half(N-2\Omega)$ and
$M_T=\half(N_p-N_n)$, where $\Omega=\half(2j+1)$ is the
half-degeneracy.
\begin{figure}
\begin{center}
\includegraphics*[width=0.75\hsize]{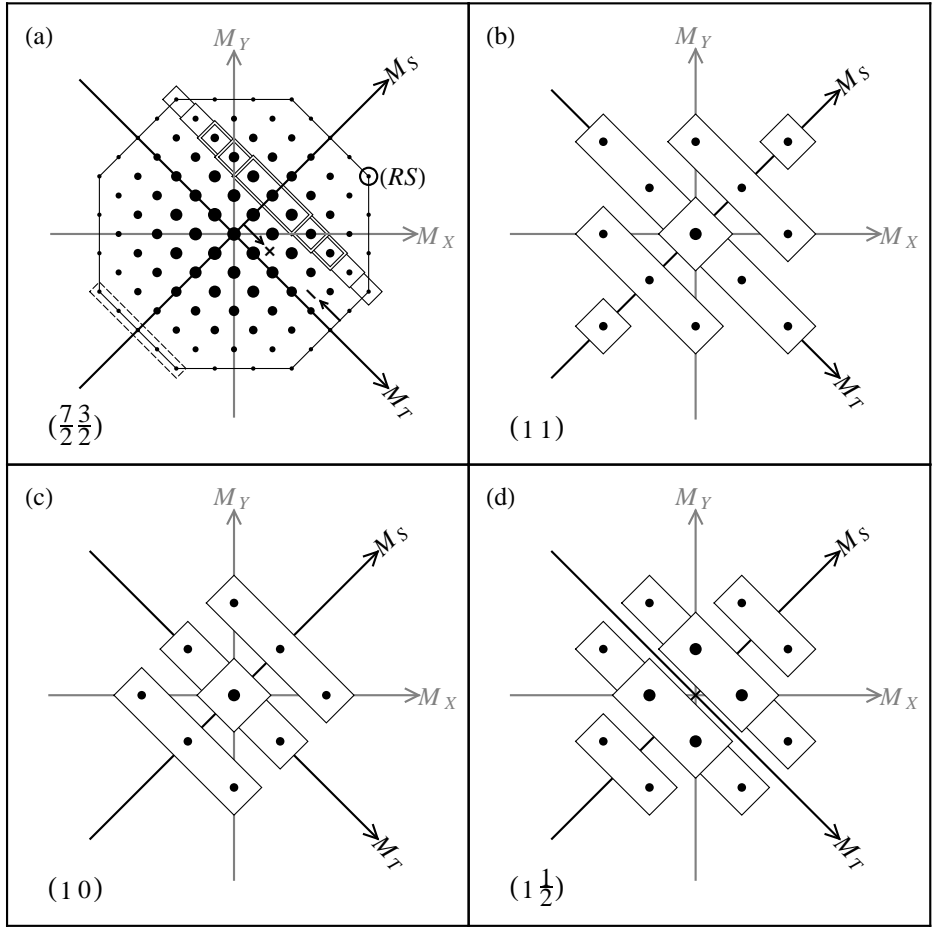}
\end{center}
\caption{Weight diagrams for $\grpso{5}$ irreps, illustrating the
decomposition into $\grpu[N]{1}\otimes\grpso[T]{3}$ irreps, {\it
i.e.}, the chain~(\chain{II}) branching.  The area of each dot
indicates the multiplicity of the weight point, and the rectangles indicate
grouping of weight points into isospin multiplets.  (a)~The general
characteristics, in the presence of branching multiplicites, are
illustrated with the irrep $(\tfrac72\tfrac32)$.  The branching is
obtained by decomposing the weights at a given value of $M_S$ into
multiplets, {\it e.g.}, for $M_S=+2$, in this example, $T=1^2$, $2^2$,
$3^2$, $4$, and $5$, where the exponents indicate branching multiplicities.  The
mathematical labeling schemes for the $\grpso{5}$ irrep are based on
the highest weight point (circle), but the $(v,t)$ labeling scheme for
pairing applications is based on the quantum numbers of the ``pair
vacuum'' $\grpu[N]{1}\otimes\grpso[T]{3}$ irrep (dashed box).  The
arrows indicate the action of the isospin laddering operators $T_\pm$
considered in the algorithms presented in the text.  (b)~Decomposition
of the low-dimensional symmetric irrep $(11)$.  (c)~Decomposition for
the low-dimensional antisymmetric irrep $(10)$, which is the adjoint
(or generator) irrep.  (d)~Decomposition of the low-dimensional
generic irrep $(1\half)$, illustrating isospin multiplets of half-integer
isospin.  }
\label{fig-weights-iso}
\end{figure}

The $\grpso{5}$ irrep $(RS)$ may be decomposed into
$\grpu[N]{1}\otimes\grpso[T]{3}$ irreps labeled by $M_S$ and $T$, as
indicated in~(\ref{eqn-so5-chains}).  By inspection of the weight
diagram [Fig.~\ref{fig-weights-iso}(a)] and considering the relation
between canonical weights (horizontal and vertical axes) and
chain~(\chain{II}) weights (diagonal axes), it can be seen that the
labels are taken from the possible values $M_S=-(R+S)$, $\ldots$,
$R+S-1$, $R+S$ and $T=0$ or $\half$, $\ldots$, $R+S-1$, $R+S$.  For a
given value of $M_S$, the highest weight $M_T$ and thus highest isospin is given by
$T_\text{max}(RS;M_S)=R+S$ for $\abs{M_S}\leq R-S$, and
$T_\text{max}(RS;M_S)=2R-\abs{M_S}$ for $\abs{M_S}>R-S$.  Note that
one of the $\grpu[N]{1}\otimes\grpso[T]{3}$ irreps, at the bottom left
of the weight diagram, is a ``pair vacuum'', annihilated by $X_-$,
$Y_-$, and $S_-$ [dashed box in Fig.~\ref{fig-weights-iso}(a)].  In
applications to proton-neutron pairing, $\grpso{5}$ irreps are
conventionally labeled not by the usual mathematical labels
(Table~\ref{tab-so5-label}), but rather by the seniority $v$ and
reduced isospin $t$.\cite{flowers1952:jj-coupling-part1}  These are
the occupation number and isospin of the pair vacuum, {\it i.e.},
$M_{S\,\text{vac}}\equiv\half(v-2\Omega)=-(R+S)$ and
$T_\text{vac}\equiv t=R-S$.

The isospin content may be seen by decomposing the weights $M_T$ for each
given value of $M_S$, {\it i.e.}, occuring on single diagonal of the
weight diagram, into isospin multiplets [solid boxes in
Fig.~\ref{fig-weights-iso}(a)].  A given pair of
$\grpu[N]{1}\otimes\grpso[T]{3}$ labels may occur more than once
within an $\grpso{5}$ irrep and is therefore labeled by a multiplicity
index $\kappa=1$, $2$, $\ldots$, $\mult(RS;M_ST)$.  The algorithm for
constructing the branching
$\grpso{5}\rightarrow\grpu[N]{1}\otimes\grpso[T]{3}$ is derived in,
{\it e.g.}, Ref.~\onlinecite{ginocchio1965:so5-quasispin}.  Simple
counting arguments then give a closed form multiplicity formula
\begin{equation}
\label{eqn-chain-II-mult}
\mult(RS;M_ST)=\begin{cases}
f(2S,T;M_S)&T\leq R-S\\
f(R+S-T,R-S;M_S)&T> R-S,
               \end{cases}
\end{equation}
where $f(u,v;w)=\lfloor\min[u,\half(u+v+w)]\rfloor
-\lceil\max[0,\half(u-v+w)]\rceil+1$.\cite{fn-f-count}  This
relation fully defines the branching rule for chain~(\chain{II}).  The
branching into $\grpu[N]{1}\otimes\grpso[T]{3}$
irreps is depicted in
Fig.~\ref{fig-weights-iso}, for example irreps of $\grpso{5}$: symmetric
[Fig.~\ref{fig-weights-iso}(b)], antisymmetric
[Fig.~\ref{fig-weights-iso}(c)], and generic
[Fig.~\ref{fig-weights-iso}(d)].

Basis states for chain~(\chain{II}) involve only a linear combination
of canonical chain~(\chain{I'}) basis states at the {\it same} point
in weight space.  This point may be labeled interchangably either
by $M_X$ and $M_Y$ or by $M_S$ and $M_T$.  Thus,
\begin{equation}
\label{eqn-chainII-bracket}
\ket[3]{(RS)\\M_{S} \kappa T\\M_{T}}=
\sum_{(XY)}
\overlap[3]{(RS)\\(XY)\\M_{X}M_{Y}}{(RS)\\M_{S} \kappa T\\M_{T}}
\,
\ket[3]{(RS)\\(XY)\\M_{X}M_{Y}},
\end{equation}
where it is to be understood that $M_X=\tfrac12(M_S+M_T)$ and
$M_Y=\tfrac12(M_S-M_T)$.  The $\grpso{4}$ irreps $(XY)$ of the
canonical basis states contributing to a given chain~(\chain{II})
basis state are constrained by the branching condition
$(RS)\rightarrow(XY)$ and by the usual angular momentum projection
rules ($\abs{M_X}\leq X$ and $\abs{M_Y}\leq Y$).

The transformation brackets in~(\ref{eqn-chainII-bracket}) are only
known in closed form for a restricted set of cases involving low
branching multiplicites.\cite{hecht1967:so5-shell-wigner}  However,
the transformation brackets can be systematically calculated for any
$\grpso{5}$ irrep, regardless of multiplicity, through two possible
procedures, outlined here, involving isospin laddering operations and
either matrix diagonalization or orthogonalization.

The first method relies upon the construction of chain~(\chain{II})
basis states as eigenstates of $\vec{T}^2$.  Suppose $\vec{T}^2$ is
written as its matrix realization in the~(\chain{I'}) basis.  The
transformation brackets appearing in~(\ref{eqn-chainII-bracket}) are
the coefficients for decomposition of the eigenstate with respect to
the chain~(\chain{I'}) basis.  Therefore, they must constitute the
entries of an eigenvector of the $\vec{T}^2$ matrix, of eigenvalue
$T(T+1)$, and may be computed by diagonalization of this matrix.
Since each chain~(\chain{II}) basis vector involves only a single weight
point, the diagonalization can be carried out separately on subspaces
corresponding to different weight points, in which case the eigenvalue
$T(T+1)$ occurs with degeneracy given by
$\mult(RS;M_ST)$ from~(\ref{eqn-chain-II-mult}).

Note that the matrix realization of $\vec{T}^2$ at weight point
$(M_XM_Y)$ is readily obtained from the known matrix
elements~(\ref{eqn-T-rme}) of the $\grpso{5}$ generators.
From~(\ref{eqn-sot3-reln}),
$\vec{T}^2=(X_0-Y_0)^2+2(\Tpm\Tmp+\Tmp\Tpm)$.  This operator can be
reexpressed in terms of spherical bitensor coupled products of the
generator $\Tss$ as $\vec{T}^2=(X_0-Y_0)^2+2[(T\times
T)^{(11)}_{00}-(T\times T)^{(00)}_{00}]$.  The requisite matrix
elements are therefore
\begin{multline}
\label{eqn-Tsqr-rme}
\me[3]{(RS)\\(X'Y')\\M_{X}M_{Y}}{\vec{T}^2}{(RS)\\(XY)\\M_{X}M_{Y}}
=
(M_X-M_Y)^2
+2\cg{(XY)}{M_XM_Y}{(11)}{00}{(X'Y')}{M_XM_Y}
\\\times
\rme[2]{(RS)\\(X'Y')}{(T\times T)^{(11)}}{(RS)\\(XY)}
-2
\rme[2]{(RS)\\(X'Y')}{(T\times T)^{(00)}}{(RS)\\(XY)}.
\end{multline}
These may be evaluated using Racah's
reduction formula,\cite{edmonds1960:am}
as naturally extended to spherical bitensors, {\it i.e.}, $\grpso{4}$
tensors, giving
\begin{multline}
\label{eqn-T11-rme}
\rme[2]{(RS)\\(X'Y')}{(T\times T)^{(11)}}{(RS)\\(XY)}
=
\sum_{(X''Y'')}
\usixj{(\half\half)}{(\half\half)}{(11)}{(XY)}{(X'Y')}{(X''Y'')}
\\\times
\rme[2]{(RS)\\(X'Y')}{T}{(RS)\\(X''Y'')}
\rme[2]{(RS)\\(X''Y'')}{T}{(RS)\\(XY)}
\end{multline}
and
\begin{equation}
\label{eqn-T00-rme}
\rme[2]{(RS)\\(X'Y')}{(T\times T)^{(00)}}{(RS)\\(XY)}
=
-\frac12 \delta_{(XY)(X'Y')} \sum_{(X''Y'')}
\rme[2]{(RS)\\(XY)}{T}{(RS)\\(X''Y'')}^2.
\end{equation}
These expressions involve only $\grpso{4}$ coupling
coefficients~(\ref{eqn-so4-coupling}), $\grpso{4}$ recoupling
coefficients~(\ref{eqn-so4-recoupling}), and
$\grpso{5}\supset\grpso{4}$ reduced matrix elements~(\ref{eqn-T-rme}),
all of which are readily calculated.

However, it is important to note that the transformation brackets must
be consistent among weight points within an isospin multiplet, since
these are connected by the isospin laddering operators $T_\pm$
[Fig.~\ref{fig-weights-iso}(a)].  The chain~(\chain{II}) basis states
of a given $M_S$, $\kappa$, and $T$ but different $M_T$ must be
related under laddering by $T_\pm$ with the correct (positive) phase.
Furthermore, in the presence of branching multiplicities, the choice
of basis vectors (resolution of the 
multiplicity) must be consistent between weight points, {\it i.e.},
laddering should not connect different $\kappa$ values at adjacent
weight points.  The relation among transformation brackets at adjacent
weight points along a diagonal of given $M_S$, obtained by comparing
the action of $T_\pm$ on the chain~(\chain{I'}) and chain~(\chain{II})
states, is
\begin{multline}
\label{eqn-iso-ladder}
[(T\mp M_T)(T\pm
M_T+1)]^{1/2}\overlap[3]{(RS)\\(X'Y')\\M_X'M_Y'}{(RS)\\M_S\kappa
T\\M_T'}
\\
=2\sum_{(XY)}
\me[3]{(RS)\\(X'Y')\\M_X'M_Y'}{T_{\pm\mp}}{(RS)\\(XY)\\M_XM_Y}
\,
\overlap[3]{(RS)\\(XY)\\M_XM_Y}{(RS)\\M_S\kappa T\\M_T},
\end{multline}
where $M_X'=M_X\pm\half$, $M_Y'=M_Y\mp\half$, and $M_T'=M_T\pm1$, and
where it is understood that $M_X=\tfrac12(M_S+M_T)$,
$M_Y=\tfrac12(M_S-M_T)$, $M_X'=\tfrac12(M_S+M_T')$, and
$M_Y'=\tfrac12(M_S-M_T')$.  At most four $\grpso{4}$ irreps $(XY)$
contribute to the sum, under the triangularity condition
$(XY)\otimes(\half\half)\rightarrow(X'Y')$.  The matrix element of
$T_{\pm\mp}$ can be calculated from~(\ref{eqn-we})
and~(\ref{eqn-T-rme}).

If $\vec{T}^2$
were diagonalized independently at each weight point, this would not
guarantee consistency between weight points under laddering.  As usual
in diagonalization problems: (1)~Eigenvectors, even in the absence of
degenerate eigenvalues, are only defined to within an arbitrary phase
(sign).  (2)~In the presence of degenerate eigenvalues (here,
multiple occurence of the same isospin at given $M_S$), the choice of
orthogonal basis vectors for the degenerate eigenspace is arbitrary.
Diagonalization of $\vec{T}^2$ must therefore
be augmented by further conditions.

A consistent prescription for the transformation brackets is provided
by diagonalizing $\vec{T}^2$ only {\it once}, for each $M_S$, at the
most central weight point on the diagonal
[Fig.~\ref{fig-weights-iso}(a)], either $M_T=0$ for integer isospin
[Fig.~\ref{fig-weights-iso}(b,c)] or $M_T=+\half$ for half-integer
isospin [Fig.~\ref{fig-weights-iso}(d)].  The remaining transformation
brackets are obtained by laddering outward, to more peripheral
weight points of larger $\abs{M_T}$ [the ``$+$'' arrow in
Fig.~\ref{fig-weights-iso}(a)].  If the transformation
brackets~(\ref{eqn-chainII-bracket}) are considered as entries of
numerical eigenvectors, then the laddering
operation~(\ref{eqn-iso-ladder}) consists of multiplication by the
(generally nonsquare and sparse) matrix realization of $T_{\pm\mp}$
between weight points.  For a vector of isospin $T$, laddering past
the weight point $M_T=T$ gives a null result.  The process terminates
at $M_T= T_\text{max}(RS;M_S)$, or $M_T= -T_\text{max}(RS;M_S)$ for
laddering towards negative $M_T$.  Although the laddering procedure
provides a {\it consistent} set of coupling coefficients, the
resolution of multiplicities arising from the
diagonalization (at $M_T=0$ or $+\half$) remains arbitrary and thus
still does not provide a {\it unique} and reproducible
prescription.  Uniqueness (to within
phase) can be obtained by further diagonalizing a ``second'' operator,
as detailed in Ref.~\onlinecite{hecht1967:so5-shell-wigner}.\cite{fn-ms-conj}

The second, alternative approach to constructing the transformation
brackets does not involve explicitly diagonalizing $\vec{T}^2$.
Rather, it is based on laddering inward along a diagonal of constant
$M_S$, from the most peripheral weight point [the ``$-$'' arrow in
Fig.~\ref{fig-weights-iso}(a)] in conjunction with orthogonalization
at each weight point.  The weight point $M_T= T_\text{max}(RS;M_S)$
contains only a single basis state, which therefore is also the
$T=T_\text{max}$ basis vector for chain~(\chain{II}), to within sign.
We are free to choose this sign {\it e.g.}, as always positive.
Laddering inward to $M_T=T_\text{max}-1$ yields the $T=T_\text{max}$
basis vector at this weight point.  If the number of basis
states ({\it i.e.}, the dimension of the subspace) at this point is
larger than one, then the remaining orthogonal vector (or vectors)
needed to span the space must be the chain~(\chain{II}) basis vector
(or vectors) of isospin $T=T_\text{max}-1$.  Ths degeneracy will be
$\mult(RS;M_S,T_\text{max}-1)$.  The orthogonal basis of good isospin
may therefore be found by Gram-Schmidt orthogonalization of a complete
but nonorthogonal basis, starting from the {\it known}
$T=T_\text{max}$ basis vector, which is supplemented by
$\mult(RS;M_S,T_\text{max}-1)$ further linearly independent vectors as
needed to span the subspace.  A {\it unique} set of transformation
brackets, both in terms of phases and resolution of multiplicities, is
obtained if a well-defined prescription is used for specifying these
further independent vectors.  For instance, one might use basis states
taken from chain~(\chain{I'}), in order of increasing weight for the
$\grpso{4}$ label $(XY)$, starting with the lowest-weight $\grpso{4}$
label available at the weight point.  (Alternatively, uniqueness can
be enforced by diagonalizing a ``second'' operator within the
$T=T_\text{max}-1$ space, supplemented by a phase convention.)  The
process of laddering followed by orthonormalization must then be
repeated for $M_T=T_\text{max}-2$, $T_\text{max}-3$, {\it etc.}, until
$M_T=0$ or $+\half$ is reached.

For large-scale computations, the choice between the two methods,
(1)~{\it diagonalization} of $\vec{T}^2$ followed by {\it outward laddering} or
(2)~{\it inward laddering} alternating with {\it orthogonalization},
will be dictated by considerations of numerical efficiency and
accuracy.  These will depend upon the numerical linear algebra
algorithms being used.  The latter method provides the simplest route
to a unique, reproducible set of phases and resolution of the branching
multiplicity.

Once the transformation brackets between bases reducing
chains~(\chain{I'}) and~(\chain{II}) have been obtained, the
transformation of reduced coupling coefficients follows immediately.
The full (unreduced) coupling coefficient may be interpreted as
the inner product of a coupled state with an uncoupled product
of two states, each described by~(\ref{eqn-chainII-bracket}).  Then,
for the
$\grpso{5}\supset\grpu[N]{1}\otimes\grpso[T]{3}$ reduced
coupling coefficient it follows from the factorization lemma and
orthonormality that
\begin{multline}
\label{eqn-chainII-transform}
\underbrace{
\cg{(R_1S_1)}{M_{S 1} \kappa_1 T_1}{(R_2S_2)}{M_{S 2} \kappa_2 T_2}{\rho(RS)}{M_{S} \kappa T}
}_{\grpso{5}\supset\grpu[N]{1}\otimes\grpso[T]{3}}
\\
=
\sum_{\substack{(X_1Y_1)(X_2Y_2)(XY)\\M_{T 1}(M_{T 2})}}
\underbrace{
\cg{T_1}{M_{T 1}}{T_2}{M_{T 2}}{T}{M_{T}}
}_{\grpso{3}}
\underbrace{
\cg{(X_1Y_1)}{M_{X 1}M_{Y 1}}{(X_2Y_2)}{M_{X 2}M_{Y 2}}{(XY)}{M_{X}M_{Y}}
}_{\grpso{4}}
\\\times
\underbrace{
\overlap[3]{(R_1S_1)\\M_{S 1} \kappa_1 T_1\\M_{T 1}}{(R_1S_1)\\(X_1Y_1)\\M_{X 1}M_{Y 1}}
\overlap[3]{(R_2S_2)\\M_{S 2} \kappa_2 T_2\\M_{T 2}}{(R_2S_2)\\(X_2Y_2)\\M_{X 2}M_{Y 2}}
\overlap[3]{(RS)\\M_{S} \kappa T\\M_{T}}{(RS)\\(XY)\\M_{X}M_{Y}}
}_{\grpso{5}\supset[\grpu[N]{1}\otimes\grpso[T]{3}\leftrightarrow\grpso{4}]}
\\\times
\underbrace{
\cg{(R_1S_1)}{(X_1Y_1)}{(R_2S_2)}{(X_2Y_2)}{\rho(RS)}{(XY)}
}_{\grpso{5}\supset\grpso{4}}
,
\end{multline}
for any value of $M_T$ allowed given isospin $T$, where again it is to
be understood that $M_X=\tfrac12(M_S+M_T)$, $M_Y=\tfrac12(M_S-M_T)$,
and similarly for $M_{X1}$, $M_{Y1}$, $M_{X2}$, and $M_{Y2}$.  The sum
over $\grpso{4}$ irrep labels is subject to the usual branching and
triangularity constraints on the canonical reduced coupling
coefficients.  For the resulting chain~(\chain{II}) reduced coupling coefficient to be
nonvanishing, it must obey the
$\grpso{5}\rightarrow\grpu[N]{1}\otimes\grpso[T]{3}$ branching
rules~(\ref{eqn-chain-II-mult}), the $\grpu[N]{1}$ additivity
condition $M_{S1}+M_{S2}=M_{S}$, and the $\grpso[T]{3}$ triangularity
condition $T_1\otimes T_2\rightarrow T$.

In Appendix~\ref{app-canonical}, the canonical reduced coupling
coefficients for the $\grpso{5}$ coupling
$(10)\otimes(1\half)\rightarrow(1\half)$ were calculated
[Fig.~\ref{fig-system-large}(c)], as a relatively simple numerical
example involving an outer multiplicity ($D=2$).  For a concrete
illustration of the transformation procedure just described, let us
consider the transformation of these coefficients into
chain~(\chain{II}) reduced coupling coefficients.  The decomposition
of the irreps $(10)$ and $(1\half)$ into
$\grpu[N]{1}\otimes\grpso[T]{3}$ irreps is shown in
Fig.~\ref{fig-weights-iso}(c,d).  For the transformation brackets, we
follow the second method above.  For instance, consider the
$M_S=+\half$ diagonal of the weight diagram for the $(1\half)$ irrep
of $\grpso{5}$ [Fig.~\ref{fig-weights-iso}(d)].  The $T=\tfrac32$
seed state at $M_T=+\tfrac32$ is
$\ket{{+\tfrac12}\tfrac32{+\tfrac32}}=+\ket{(1\tfrac12){+1}{-\tfrac12}}$, where
we abbreviate chain~(\chain{I'}) basis states as
$\ket{(XY)M_XM_Y}$ and chain~(\chain{II}) basis states as
$\ket{M_STM_T}$.  Laddering inward to $M_T=+\tfrac12$ 
yields $T=\tfrac32$ state
\begin{displaymath}
\ket{{+\tfrac12}\tfrac32{+\tfrac12}}=
+\sqrt{\tfrac56}\ket{(\tfrac12 0){+\tfrac12}0}
+\sqrt{\tfrac16}\ket{(\tfrac12 1){+\tfrac12}0}.
\end{displaymath}
Orthogonalization, using $\ket{(\tfrac12 0){+\tfrac12}0}$ as the
independent Gram-Schmidt basis vector, gives $T=\tfrac12$ state
\begin{displaymath}
\ket{{+\tfrac12}\tfrac12{+\tfrac12}}=
+\sqrt{\tfrac16}\ket{(\tfrac12 0){+\tfrac12}0}
-\sqrt{\tfrac56}\ket{(\tfrac12 1){+\tfrac12}0}.
\end{displaymath}
Continued laddering to negative $M_T$ gives the remaining
transformation brackets along
the diagonal.  
Straightforward application
of~(\ref{eqn-chainII-transform}), using the full set of transformation
brackets derived in this fashion, then yields the chain~(\chain{II})
reduced coupling coefficients in Table~\ref{tab-chainII-coeffs}.  Note
that the coefficients are either symmetric or antisymmetric under the $\grpu[N]{1}$
particle-hole conjugation operation $M_S\rightarrow
-M_S$ (see Ref.~\onlinecite{hecht1967:so5-shell-wigner}).
\begin{table}
\caption{Chain~(\chain{II}) reduced coupling coefficients for the $\grpso{5}$ coupling
$(10)\otimes(1\half)\rightarrow(1\half)$, obtained from the canonical
coupling coefficients of Fig.~\ref{fig-system-large}(c) according to
the transformation conventions described in the text.  For each
$\grpu[N]{1}\otimes\grpso[T]{3}$ coupling
$(M_{S\,1}T_1)\otimes(M_{S\,2}T_2)\rightarrow (M_{S}T)$, the
coefficients are given for $\grpso{5}$ outer multiplicity index
values $\rho=1$ and $2$, respectively.
}
\label{tab-chainII-coeffs}
\input{racah_tab03.tex}
\end{table}


The $\grpso[L]{3}$ subalgebra of chain~(\chain{III}) is the maximal
$\grpso{3}$ subalgebra, {\it i.e.}, one which is not contained within
any larger proper subalgebra of $\grpso{5}$.  This subalgebra is
obtained by
letting\cite{corrigan1976:oscillator,rowe1994:so5-so4-vcs,debaerdemacker2007:so5-cartan}
\begin{equation}
\label{eqn-chainIII-gen}
\begin{gathered}
L^{(1)}_{+1}= -\sqrt2X_+-\sqrt6\Tmp
\quad
L^{(1)}_{0}= X_0+3Y_0
\quad
L^{(1)}_{-1}= \sqrt2X_-+\sqrt6\Tpm
\\
O^{(3)}_{+3}= -\sqrt5Y_+
\quad
O^{(3)}_{+2}= \sqrt{10}\Tpp
\quad
O^{(3)}_{+1}= -\sqrt3X_++2\Tmp
\quad
O^{(3)}_{0}= 3X_0-Y_0
\\
O^{(3)}_{-1}= \sqrt3X_--2\Tpm
\quad
O^{(3)}_{-2}= -\sqrt{10}\Tmm
\quad
O^{(3)}_{-3}= \sqrt5Y_-.
\end{gathered}
\end{equation}
Then $\grpso[L]{3}$ has
generators $L^{(1)}_{M_L}$, where, as usual,
$L^{(1)}_{\pm1}=\mp\tfrac1{\sqrt2}L_\pm$.  The remaining generators
$O^{(3)}_{M_L}$ constitute an octupole tensor with respect to
$\grpso[L]{3}$.  The commutation relations of the generators are given
(most compactly in spherical tensor coupled
form\cite{french1966:multipole}) by
\begin{equation}
\label{eqn-chainIII-comm}
\begin{aligned}
[L,L]^{(1)}&=-\sqrt2L \quad & [L,O]^{(3)}&=-2\sqrt3O\\
[O,O]^{(1)}&=-2\sqrt7L & [O,O]^{(3)}&=\sqrt6O.
\end{aligned}
\end{equation}
The $\grpso[L]{3}$ weight $M_L$ is related to the canonical weights by
$M_{L}=M_{X}+3M_{Y}$, according to~(\ref{eqn-chainIII-gen}), and thus
defines an oblique axis in the weight space.  The $\grpso{5}$
generators are shown classified according to this weight in
Fig.~\ref{fig-root-chains}(d).  The dashed lines connect the canonical
generators of Sec.~\ref{sec-alg}, which have good $M_L$ but not $L$,
to the linear combinations $L^{(1)}_{M_L}$ and $O^{(3)}_{M_L}$, which
have both good $M_L$ and good $L$.

The labels for basis states reducing the $\grpso[L]{3}$ subalgebra are
indicated in~(\ref{eqn-so5-chains}).  Basis states for
chain~(\chain{III}) involve a linear combination of chain~(\chain{I'})
basis states of the same $M_L$.  The transformation between bases therefore
involves a sum not only over $(XY)$, as
in~(\ref{eqn-chainII-bracket}), but also over distinct weight points $(M_XM_Y)$,
\begin{equation}
\label{eqn-chainIII-bracket}
\ket[3]{(RS)\\ \alpha L\\M_{L}}=
\sum_{\substack{(XY)\\M_X(M_Y)}}
\overlap[3]{(RS)\\(XY)\\M_{X}M_{Y}}{(RS)\\ \alpha L\\M_{L}}
\,
\ket[3]{(RS)\\(XY)\\M_{X}M_{Y}},
\end{equation}
subject to the constraint $M_{L}=M_{X}+3M_{Y}$.  The transformation
brackets may be systematically evaluated using an adaptation of either
of the methods proposed above for Chain~(\chain{II}):
(1)~diagonalization of $\vec{L}^2$ for the $M_L=0$ or $+\half$
subspace of the $\grpso{5}$ irrep, followed by laddering outward to
larger-$M_L$ spaces, or (2)~laddering inward from the maximal-$M_L$
space, in conjunction with orthogonalization within each successive
lower-$M_L$ space.  For a unique resolution of the
$\grpso{5}\supset\grpso[L]{3}$ branching multiplicity, it has been
suggested that chain~(\chain{III}) basis states be chosen in which the
octupole tensor is diagonal, equivalent to diagonalizing the Hermitian
``second'' operator $(L\times L \times O \times L)^{(0)}+ (L\times O
\times L \times L)^{(0)}$.\cite{rowe1995:k-canonical}
Once transformation brackets have been obtained, the transformation of
reduced coupling coefficients is given by
\begin{multline}
\label{eqn-chainIII-transform}
\underbrace{
\cg{(R_1S_1)}{\alpha_1 L_1}{(R_2S_2)}{\alpha_2 L_2}{\rho(RS)}{\alpha L}
}_{\grpso{5}\supset\grpso[L]{3}}
\\
=
\sum_{\substack{(X_1Y_1)(X_2Y_2)(XY)\\M_{X 1}(M_{Y 1})M_{X 2}(M_{Y 2})M_{X}(M_{Y})\\M_{L 1}(M_{L 2})}}
\underbrace{
\cg{L_1}{M_{L 1}}{L_2}{M_{L 2}}{L}{M_{L}}
}_{\grpso{3}}
\underbrace{
\cg{(X_1Y_1)}{M_{X 1}M_{Y 1}}{(X_2Y_2)}{M_{X 2}M_{Y 2}}{(XY)}{M_{X}M_{Y}}
}_{\grpso{4}}
\\\times
\underbrace{
\overlap[3]{(R_1S_1)\\ \alpha_1 L_1\\M_{L 1}}{(R_1S_1)\\(X_1Y_1)\\M_{X 1}M_{Y 1}}
\overlap[3]{(R_2S_2)\\ \alpha_2 L_2\\M_{L 2}}{(R_2S_2)\\(X_2Y_2)\\M_{X 2}M_{Y 2}}
\overlap[3]{(RS)\\ \alpha L\\M_{L}}{(RS)\\(XY)\\M_{X}M_{Y}}
}_{\grpso{5}\supset[\grpso[L]{3}\leftrightarrow\grpso{4}]}
\\\times
\underbrace{
\cg{(R_1S_1)}{(X_1Y_1)}{(R_2S_2)}{(X_2Y_2)}{\rho(RS)}{(XY)}
}_{\grpso{5}\supset\grpso{4}}
,
\end{multline}
for any value of $M_L$ allowed given angular momentum $L$, where the
summations over $M_{X1}$, $M_{Y1}$, $M_{X2}$, $M_{Y2}$, $M_X$, and
$M_Y$ are subject to the constraints $M_{L 1}=M_{X 1}+3M_{Y 1}$, $M_{L
2}=M_{X 2}+3M_{Y 2}$, and $M_{L}=M_{X}+3M_{Y}$.

\vfil


\input{racah.bbl}

\end{document}

%% file: racah_tab01.tex
\begin{minipage}{\hsize} 
\begin{center}
\begin{ruledtabular}
\begin{tabular}{c|cccc}
& $\Tpp$& $\Tpm$& $\Tmp$& $\Tmm$\\
\hline
$\Tpp$& $0$& $-\tfrac1{\sqrt2}X_{+1}$& $-\tfrac1{\sqrt2}Y_{+1}$&  $-\tfrac12(X_0+Y_0)$\\
$\Tpm$& $+\tfrac1{\sqrt2}X_{+1}$& $0$&   $+\tfrac12(X_0-Y_0)$& $-\tfrac1{\sqrt2}Y_{-1}$ \\
$\Tmp$& $+\tfrac1{\sqrt2}Y_{+1}$& $-\tfrac12(X_0-Y_0)$& $0$& $-\tfrac1{\sqrt2}X_{-1}$ \\
$\Tmm$& $+\tfrac12(X_0+Y_0)$& $+\tfrac1{\sqrt2}Y_{-1}$&  $+\tfrac1{\sqrt2}X_{-1}$& $0$
\end{tabular}
\end{ruledtabular}
\end{center}
\end{minipage}

%% file: racah_tab02.tex
\begin{ruledtabular}
\begin{tabular}{@{\extracolsep{0pt}}l@{\extracolsep{\fill}}r@{\extracolsep{0pt}}l@{\extracolsep{\fill}}r@{\extracolsep{0pt}}l@{\extracolsep{\fill}}l@{\extracolsep{\fill}}l}
Labels &  \multicolumn{2}{l}{Range} & \multicolumn{2}{l}{Relations} & Description & Refs.
\\
\hline
$[l_1l_2]$&                 $l_1$&$=l_2,l_2+1,\ldots$&           &---&         $\grpso{5}$ Cartan highest weight & \onlinecite{flowers1964:quasispin,kemmer1968:so5-irreps-1}\\
          &                 $l_2$&$=0,\tfrac12,\ldots$&\\
$(a_1a_2)$&                 $a_1$&$=0,1,\ldots$&            $a_1$&$=l_1-l_2$&  $\grpso{5}$ Dynkin & \onlinecite{speiser1964:lie-compact}\\
          &                 $a_2$&$=0,1,\ldots$&            $a_2$&$=2l_2$\\
$(vf)$&                     $v$&$=0,1,\ldots$&              $v$&$=l_1-l_2$&  $\grpso{5}$ Dynkin (modified) & \onlinecite{turner2006:so5-so3-vcs}\\
          &                 $f$&$=0,\tfrac12,\ldots$&       $f$&$=l_2$\\
$\langle l_1' l_2'\rangle$& $l_1'$&$=l_2',l_2'+1,\ldots$&   $l_1'$&$=l_1+l_2$&  $\grpsp{4}$ Cartan highest weight & \onlinecite{flowers1952:jj-coupling-part1,helmers1961:shell-sp} \\
                          & $l_2'$&$=0,1,\ldots$&           $l_2'$&$=l_1-l_2$  \\
$(a_1'a_2')$&               $a_1'$&$=0,1,\ldots$&           $a_1'$&$=2l_2$&    $\grpsp{4}$ Dynkin & \onlinecite{behrends1962:groups-strong,ginocchio1965:so5-quasispin}\\
                          & $a_2'$&$=0,1,\ldots$&           $a_2'$&$=l_1-l_2$\\
$(RS)$&                     $R$&$=S,S+\tfrac12,\ldots$&     $R$&$=\tfrac12(l_1+l_2)$&  $\grpso{3}\otimes\grpso{3}$ highest weight & \onlinecite{hecht1965:so5-wigner,parikh1965:isospin-seniority}\\
                          & $S$&$=0,\tfrac12,\ldots$&       $S$&$=\tfrac12(l_1-l_2)$\\
\end{tabular}
\end{ruledtabular}

%% file: racah_tab03.tex
\begin{ruledtabular}
\begin{tabular}{rrrrrrrrrrrrrrrr}
$M_{S\,1}$&$M_{S\,2}$&$M_{S}$&$T_1$&$T_2$&$T$&\multicolumn{2}{c}{Coefficients}&$M_{S\,1}$&$M_{S\,2}$&$M_{S}$&$T_1$&$T_2$&$T$&\multicolumn{2}{c}{Coefficients}\\
\hline
$1$&$\tfrac{1}{2}$&$\tfrac{3}{2}$&$1$&$\tfrac{1}{2}$&$\tfrac{1}{2}$&$\sqrt{\tfrac{1}{3}}$&$-\sqrt{\tfrac{1}{7}}$&$0$&$-\tfrac{1}{2}$&$-\tfrac{1}{2}$&$0$&$\tfrac{1}{2}$&$\tfrac{1}{2}$&$\sqrt{\tfrac{1}{30}}$&$\sqrt{\tfrac{9}{70}}$\\
&&&$1$&$\tfrac{3}{2}$&$\tfrac{1}{2}$&$\sqrt{\tfrac{4}{15}}$&$\sqrt{\tfrac{16}{35}}$&&&&$0$&$\tfrac{3}{2}$&$\tfrac{3}{2}$&$\sqrt{\tfrac{1}{30}}$&$-\sqrt{\tfrac{9}{70}}$\\
$1$&$-\tfrac{1}{2}$&$\tfrac{1}{2}$&$1$&$\tfrac{1}{2}$&$\tfrac{1}{2}$&$-\sqrt{\tfrac{4}{45}}$&$-\sqrt{\tfrac{12}{35}}$&&&&$1$&$\tfrac{1}{2}$&$\tfrac{1}{2}$&$-\sqrt{\tfrac{1}{10}}$&$\sqrt{\tfrac{1}{210}}$\\
&&&$1$&$\tfrac{1}{2}$&$\tfrac{3}{2}$&$\sqrt{\tfrac{2}{9}}$&$0$&&&&$1$&$\tfrac{1}{2}$&$\tfrac{3}{2}$&$0$&$-\sqrt{\tfrac{4}{21}}$\\
&&&$1$&$\tfrac{3}{2}$&$\tfrac{1}{2}$&$-\sqrt{\tfrac{4}{9}}$&$0$&&&&$1$&$\tfrac{3}{2}$&$\tfrac{1}{2}$&$0$&$\sqrt{\tfrac{8}{21}}$\\
&&&$1$&$\tfrac{3}{2}$&$\tfrac{3}{2}$&$-\sqrt{\tfrac{1}{9}}$&$\sqrt{\tfrac{3}{7}}$&&&&$1$&$\tfrac{3}{2}$&$\tfrac{3}{2}$&$-\sqrt{\tfrac{1}{2}}$&$-\sqrt{\tfrac{1}{42}}$\\
$1$&$-\tfrac{3}{2}$&$-\tfrac{1}{2}$&$1$&$\tfrac{1}{2}$&$\tfrac{1}{2}$&$-\sqrt{\tfrac{1}{3}}$&$\sqrt{\tfrac{1}{7}}$&$0$&$-\tfrac{3}{2}$&$-\tfrac{3}{2}$&$0$&$\tfrac{1}{2}$&$\tfrac{1}{2}$&$\sqrt{\tfrac{3}{10}}$&$\sqrt{\tfrac{1}{70}}$\\
&&&$1$&$\tfrac{1}{2}$&$\tfrac{3}{2}$&$\sqrt{\tfrac{2}{15}}$&$\sqrt{\tfrac{8}{35}}$&&&&$1$&$\tfrac{1}{2}$&$\tfrac{1}{2}$&$-\sqrt{\tfrac{1}{10}}$&$\sqrt{\tfrac{27}{70}}$\\
$0$&$\tfrac{3}{2}$&$\tfrac{3}{2}$&$0$&$\tfrac{1}{2}$&$\tfrac{1}{2}$&$-\sqrt{\tfrac{3}{10}}$&$-\sqrt{\tfrac{1}{70}}$&$-1$&$\tfrac{3}{2}$&$\tfrac{1}{2}$&$1$&$\tfrac{1}{2}$&$\tfrac{1}{2}$&$\sqrt{\tfrac{1}{3}}$&$-\sqrt{\tfrac{1}{7}}$\\
&&&$1$&$\tfrac{1}{2}$&$\tfrac{1}{2}$&$-\sqrt{\tfrac{1}{10}}$&$\sqrt{\tfrac{27}{70}}$&&&&$1$&$\tfrac{1}{2}$&$\tfrac{3}{2}$&$-\sqrt{\tfrac{2}{15}}$&$-\sqrt{\tfrac{8}{35}}$\\
$0$&$\tfrac{1}{2}$&$\tfrac{1}{2}$&$0$&$\tfrac{1}{2}$&$\tfrac{1}{2}$&$-\sqrt{\tfrac{1}{30}}$&$-\sqrt{\tfrac{9}{70}}$&$-1$&$\tfrac{1}{2}$&$-\tfrac{1}{2}$&$1$&$\tfrac{1}{2}$&$\tfrac{1}{2}$&$-\sqrt{\tfrac{4}{45}}$&$-\sqrt{\tfrac{12}{35}}$\\
&&&$0$&$\tfrac{3}{2}$&$\tfrac{3}{2}$&$-\sqrt{\tfrac{1}{30}}$&$\sqrt{\tfrac{9}{70}}$&&&&$1$&$\tfrac{1}{2}$&$\tfrac{3}{2}$&$\sqrt{\tfrac{2}{9}}$&$0$\\
&&&$1$&$\tfrac{1}{2}$&$\tfrac{1}{2}$&$-\sqrt{\tfrac{1}{10}}$&$\sqrt{\tfrac{1}{210}}$&&&&$1$&$\tfrac{3}{2}$&$\tfrac{1}{2}$&$-\sqrt{\tfrac{4}{9}}$&$0$\\
&&&$1$&$\tfrac{1}{2}$&$\tfrac{3}{2}$&$0$&$-\sqrt{\tfrac{4}{21}}$&&&&$1$&$\tfrac{3}{2}$&$\tfrac{3}{2}$&$-\sqrt{\tfrac{1}{9}}$&$\sqrt{\tfrac{3}{7}}$\\
&&&$1$&$\tfrac{3}{2}$&$\tfrac{1}{2}$&$0$&$\sqrt{\tfrac{8}{21}}$&$-1$&$-\tfrac{1}{2}$&$-\tfrac{3}{2}$&$1$&$\tfrac{1}{2}$&$\tfrac{1}{2}$&$-\sqrt{\tfrac{1}{3}}$&$\sqrt{\tfrac{1}{7}}$\\
&&&$1$&$\tfrac{3}{2}$&$\tfrac{3}{2}$&$-\sqrt{\tfrac{1}{2}}$&$-\sqrt{\tfrac{1}{42}}$&&&&$1$&$\tfrac{3}{2}$&$\tfrac{1}{2}$&$-\sqrt{\tfrac{4}{15}}$&$-\sqrt{\tfrac{16}{35}}$\\
\end{tabular}
\end{ruledtabular}

%% file: racah.bbl
\providecommand{\ELSEVIER}{}
\newcommand{\identity}[1]{{#1}}

%% file: racah.bbl
\begin{thebibliography}{68}
\bibitem{fn-nullmethod}
\bibinfo{note}{Euler row reduction permits solution for the null vectors in
  exact symbolic arithmetic, as illustrated in the examples of
  Appendix~\ref{app-canonical}. However, for large-scale numerical calculation
  of coupling coefficients, $QR$ decomposition or singular value decomposition
  algorithms would be more suitable.}
\bibitem{fn-weights-I-II}
\bibinfo{note}{In relation to the Cartan weight operators of chain~(\chain{I}),
  observe that $S_0=J_3$ and $T_0=N_3$. Thus, the natural weight labels of
  chain~(\chain{I}) and chain~(\chain{II}) are identical, with
  $[\lambda_1\lambda_2]=[M_SM_T]$. However, in the context
  of~(\ref{eqn-so4-gen}), $N_3$ was not the ``$M$'' weight operator of any
  $\grpso{3}$ algebra.}
\bibitem{fn-f-count}
\bibinfo{note}{The function $f(u,v;w)$ counts the number of lattice points
  $(x,y)$, with $x=-u$, $-u+2$, $\ldots$, $+u$ and $y=-v$, $-v+1$, $\ldots$,
  $+v$, such that $x+y=w$.}
\end{thebibliography}
